\documentclass{optica-article}

\journal{opticajournal} 

\articletype{Research Article}

\usepackage{lineno}
\usepackage[OT2,T1]{fontenc}
\DeclareSymbolFont{cyrletters}{OT2}{wncyr}{m}{n}
\DeclareMathSymbol{\Sha}{\mathalpha}{cyrletters}{"58}
\begin{document}

\title{Spatial coherence of synchrotron radiation degraded by grating monochromators}

\author{R. Khubbutdinov,\authormark{1,*} G. Geloni,\authormark{2} E. Saldin,\authormark{1},  and K. Bagschik\authormark{1}}
\address{\authormark{1}Deutsches Elektronen-Synchrotron DESY, Notkestraße 85, Hamburg, D-22607, Germany\\
\authormark{2}European XFEL, 4, Holzkoppel Street, 22869 Schenefeld, Germany\\
}

\email{\authormark{*}ruslan.khubbutdinov@desy.de} 




\section{Introduction}
Spatial coherence and spectral brightness play a key role in a large number of synchrotron experiments. The demand for coherent flux for these experiments has increased enormously in recent years. Synchrotron facilities worldwide have or will upgrade their storage rings to the fourth generation (4GSR) \cite{tavares2014max, weckert2015, schroer2018petra, Rodrigues2018, schroer2019petra} to meet this demand and open up new scientific opportunities. These storage rings are based on the Multi-Bend Achromat (MBA) technology and allow horizontal emittances in the order of tens of pmrad. This results in a significant increase in the spectral brightness and coherent flux of the photon sources \cite{geloni2018effects, khubbutdinov2019a} and enables new opportunities, in particular for coherent imaging and scattering experiments such as X-ray ptychography \cite{lipeng2022,dasilva2019} and XPCS \cite{remont24,narayanan23,berkowicz22}. 

The increased spectral brightness of the source poses significant challenges for the beamline optics. The photon beam transport system must preserve the high spatial (transverse) coherence of the source to make it available for the actual experiment. Previous studies have investigated the spatial coherence properties of the synchrotron sources \cite{geloni2008transverse, singer2014coherence, geloni2015brightness, geloni2018effects, khubbutdinov2019a, Vartanyants2020} and how they degrade due to various factors, such as optical surface defects \cite{surf, surf2, surf3, surf4}, incoherent scattering \cite{scatt, scatt2}, or vibrations of optical elements \cite{vibr, vibr2, bagschik2020direct}, including the Kirkpatrick-Baez (KB) system and monochromator. Depending on the number of optical components at the beamline, the effect of vibrations imposes the most significant influence on spatial coherence so far. 

In the soft X-ray region, 4GSRs provide spatial coherence of almost 100\% in both the horizontal and vertical directions. The spatial coherence reducing effects of the beamline optics are particularly strong here. Soft X-ray beamlines use grating monochromators to increase energy resolution. There are different types of grating monochromators, e.g. planar gratings with a downstream focusing element or self-focussing gratings such as the Varibale Line Spacing (VLS) grating. The energy bandwidth is defined in combination with an exit slit. Studies at the soft X-ray beamline P04 at PETRA III have shown that the vertical exit slit of the monochromator affects the spatial coherence in the vertical direction \cite{bagschik2020,khubbutdinov2022soft}. Despite this expected effect, it has been shown that the measured spatial coherence in the dispersion direction is far below the theoretically predicted value \cite{bagschik2020direct,skopintsev2014,Rose2015a}. So far, gratings have not been considered as a source of spatial coherence degradation.

In this paper we present a physical and theoretical description for the significant degradation of the spatial coherence properties of the photon beam due to the use of grating monochromators at synchrotron facilities. We present a comprehensive study of the effect of spatial coherence degradation as a function of grating parameters under different focusing and propagation conditions. Emphasis has been placed on the spatial coherence degradation effect introduced by the grating alone, thus assuming a fully coherent incident photon beam generated by a filament source. The mathematical description and the theoretical analysis of the spatial coherence degradation are presented in the framework of statistical optics.

The first sections give an overview of the synchrotron radiation pulse structure and the mathematical description of the undulator radiation, then the basic theory of statistical optics as applied to the synchrotron source. The effect of the dispersion introduced by the grating monochromator on the spatial coherence properties of the X-ray radiation is investigated in the following sections. Individual effects of free-space propagation and focusing of the photon beam with dispersion are considered, as well as the effects of aberration and photon beam clipping by the exit slit. An evaluation of the spatial coherence properties of the photon beam and the energy resolution of the monochromator are given in these sections. The study concludes with a discussion and summary of the analysis, followed by an explanation of the effect and its applicability to synchrotron radiation sources.

\section{\label{sec:undi} Statistical properties of a synchrotron radiation source}
This section describes the mathematical basis of the statistical methods used for synchrotron radiation (SR) sources. In order to understand the origin of spatial coherence degradation caused by gratings, it is necessary to consider the pulse structure and statistics of synchrotron radiation sources. Therefore, this section briefly discusses the pulse structure of synchrotron radiation and the mathematical description of the X-ray source, followed by the basic theory of coherence in the framework of statistical optics. 

\subsection{Pulse structure}\label{PulseSR}
In this paper an undulator is considered as the primary source of X-ray radiation. The source has an intrinsic stochastic structure driven by shot-noise statistics. This means that the fluctuations of the photon beam density are random in the six dimensional phase space volume containing two spatial, two angular, time and energy projections \cite{geloni2008transverse,geloni2018effects}. These fluctuations follow a Gaussian distribution. As a result, the produced radiation field has random amplitudes and phases, implying that the synchrotron radiation process is a Gaussian random process with shot-noise statistics imprinted in the radiation structure \cite{geloni2008transverse}. The latter manifests itself as longitudinal (or spectral) and transverse individual spikes in the radiation pulse. The presence of individual temporal spikes in a pulse implies the existence of certain coherent regions in the time domain with a characteristic coherence time $\tau_c$. As such, the width of these individual spikes can be roughly estimated from the simple Fourier transform (FT) theory relations 

\begin{equation} \label{undi7}
	\Delta\omega\Delta\tau=2\pi,
\end{equation}

Additionally, it was shown that the SR process is non-stationary and, consequently, non-ergodic but a quasi-stationary process \cite{geloni2008transverse}. In the framework of statistical optics considering quasi-stationary processes, one can estimate characteristic times of such radiation pulses by considering a Wiener-Khinchin theorem \cite{mandel1995}.
According to the theorem, the coherence time of the processes with Gaussian spectral density is $\tau_c$=$\sqrt{\pi}/\sigma_\omega$, where $\sigma_\omega$ is the sigma and $\Delta\omega$ = $2\sqrt{2ln(2)}$ is the width of the spectrum. The spectral width of the radiation produced by an undulator is \cite{Jackson1999a, alferov1989radiation, onuki2002}
\begin{equation} \label{undi9}
	\frac{\Delta\omega}{\omega} = \frac{1}{nN_u},
\end{equation} 
where $n$ is the harmonic number and $N_u$ is the number of magnet periods. The typical number of periods is about $N_u = 10^2$, so that the spectral bandwidth in the soft X-ray range (500 eV - 4 keV) is about $\Delta\omega\sim 10^{16}$ Hz, and the corresponding coherence time is $\tau_c\sim 10^{-16}$ s. The characteristic times of the electron bunch can reach values of $\sigma_t\sim$ 30$\cdot10^{-12}$ s. From this, it can be seen that on the scale of pulse duration there are $N$=$\sigma_t/\tau_c$= $10^5$ random intensity fluctuations or temporal spikes. 
\begin{figure}[ht!]
	\centering\includegraphics[width=1.0\textwidth]{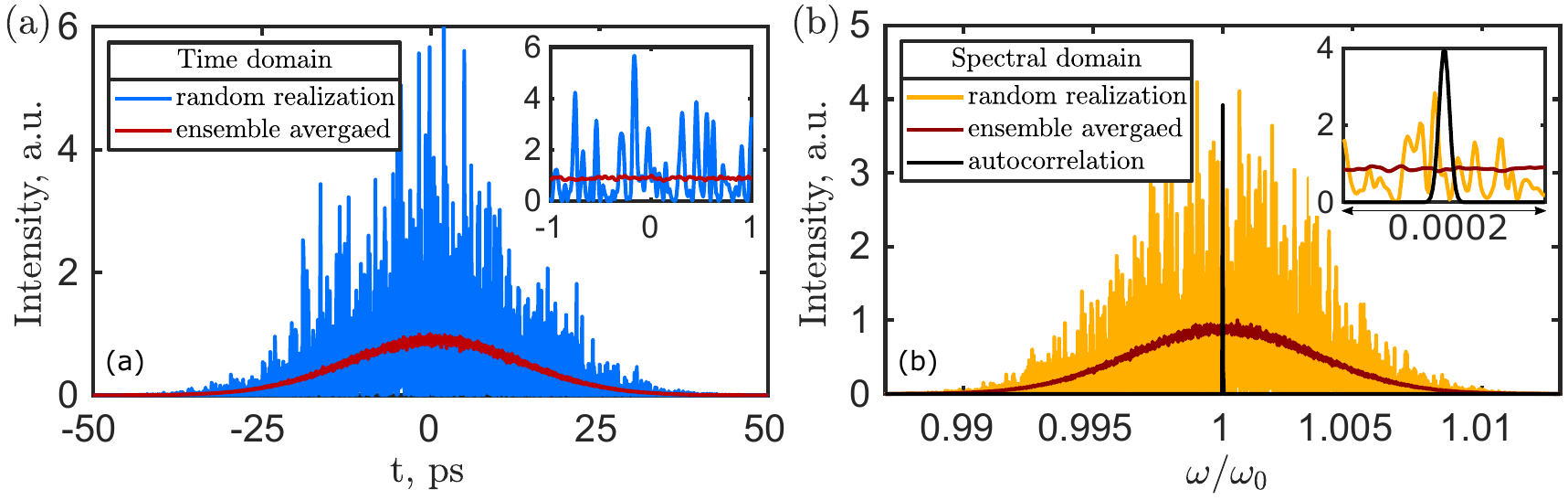}
	\caption{Characteristic undulator radiation field in the time-frequency domain that has an intrinstic stochastic structure driven by shot-noise statistics. Gaussian intensity fluctuations of the undulator radiation in time (a) and spectral domain (b) with characteristic pulse duration $T = 30\cdot10^{-12}$s and coherence time $\tau_c = 1\cdot10^{-16}$s (a), resulting in a spectral width of $\Delta\omega/\omega\sim 0.01$ and spectral spike width $\Delta\omega_c/\omega\sim 10^{-7}$ (b). The red profile corresponds to the ensemble average of a large number of random realizations (blue). The inset shows the characteristic scale of the autocorrelation function.} \label{Fluc1}
\end{figure}

Looking at the spectral domain by analogy, one can estimate a coherent spike in the spectrum $\Delta\omega_c\sim 2\pi/\sigma_t$. In this case, the value of $\Delta\omega_c$ is about $10^{11}$ Hz, so the spectrum also contains about $10^5$ spikes. This is shown in Fig. \ref{Fluc1} along with the Fourier relation. As can be seen from this analysis, on the scale of the average spectrum, the coherent region (spike width) is $\Delta\omega_c/\omega_0\approx 10^{-7}$ and rapidly disappears as one approaches the typical pulse duration of SR. Typically, today's monochromators cannot resolve a single spike, leading to the case of a convenient model for coherence analysis. 

\subsection{Mathematical description}\label{sourceM}

The mathematical tools of statistical optics are required to analyse the effect of spatial coherence degradation caused by gratings used in grating monochromators. For simplicity, the description and theoretical derivations are usually presented in the $\{\mathbf{r}$-$\omega\}$-domain, which implies a one-to-one correspondence to the $\{\mathbf{r}$-$t\}$-domain via FT relations, where the following FT pair is considered for the radiation fields $E(\mathbf{r},t)$ and $E(\mathbf{r},\omega)$ 
\begin{equation}\label{inter1a}
	E(\mathbf{r},t) = \frac{1}{2\pi} \int_{-\infty}^{\infty} E(\mathbf{r},\omega)e^{-i \omega t} d\omega,
\end{equation}
\begin{equation}\label{inter1b}
	E(\mathbf{r},\omega) = \int_{-\infty}^{\infty} E(\mathbf{r},t)e^{i \omega t} dt.
\end{equation}
Statistical processes such as undulator radiation are treated with the concept of the statistical ensemble \cite{mandel1995}, which averages over an ensemble or many realizations of the process under consideration. In this context, the averaging is performed over the distributions of the electron bunches \cite{geloni2008transverse}. 

The radiation field $E_{k\perp}(\boldsymbol{r},\omega)$ of the $k$th electron can be written as a function of the deflection angle $\boldsymbol{\eta}_k$, the offset from the undulator axis $\mathbf{l}_k$ and the offset from the electron beam energy $\gamma_k$ (energy spread) as \cite{geloni2018effects, geloni2008transverse}

\begin{equation}\label{undi2}
	\begin{aligned}
		E_{k\perp} = f(\boldsymbol{r},\omega,\boldsymbol{\eta}_k,\mathbf{l}_k,\gamma_k) 
	\end{aligned}
\end{equation} 

where $\boldsymbol{r}$=$\{x,y\}$ is the point of observation. The radiation field $E_{k\perp}$ considered in Eq. \eqref{undi2} is distributed in the transverse plane perpendicular to the electron and photon beam propagation direction at a certain distance $z$ from the undulator center (in the far zone, considering the paraxial approximation). The total radiation field is the sum of all individual electron contributions as given by

\begin{equation} \label{undi3}
	E_{\perp}(\boldsymbol{r},\omega) =  \sum_{k=1}^{N_e} E_{k\perp}(\boldsymbol{r},\omega,\boldsymbol{\eta}_k,\mathbf{l}_k,\gamma_k),
\end{equation}

where $N_e$ is the number of electrons in the beam. The parameters $\boldsymbol{\eta}_k$ and $\mathbf{l}_k$ are random variables and within the range of the electron beam divergence and size distributions. Another important random parameter $t_k$, the arrival time of the $k$th electron at the undulator entrance, must also be taken into account, as it greatly simplifies our model and the corresponding calculations when analysing spatial coherence. According to the Fourier transform in the $\{\omega,t\}$ domain in Eq. \eqref{inter1b}, applying a shift theorem to the field $E(\mathbf{r},t-t_k)$ delayed by the time $t_k$, there will be an additional factor in the transition from the $\{\mathbf{r},t\}$ to the $\{\mathbf{r},\omega\}$-domain defined by

\begin{equation}\label{undi5}
	\int_{-\infty}^{\infty} E_{\perp}(\mathbf{r},t-t_k)e^{i \omega t} dt = E_{\perp}(\mathbf{r},\omega)e^{i\omega t_k}.
\end{equation}

The total radiation field $E(\boldsymbol{r},\omega)$ that accounts for all possible random shifts within the electron bunch is the sum of the partial contributions as given by \cite{geloni2008transverse}

\begin{equation} \label{undi6}
	E(\boldsymbol{r},\omega) =  \sum_{k=1}^{N_e} E_{k\perp}(\boldsymbol{r},\omega,\boldsymbol{\eta}_k,\mathbf{l}_k,\gamma_k)e^{i\omega t_k},
\end{equation}

where $t_k$ is in the range of the electron bunch duration $\sigma_t$. It is also assumed that the random arrival times $t_k$ are independent of the random transverse shifts $\mathbf{l}_k$ and $\boldsymbol{\eta}_k$, which is the case at synchrotron facilities. In general $\boldsymbol{\eta}_k$=$\{\eta_x,\eta_y\}_k$, $\mathbf{l}_k$=$\{l_x,l_y\}_k$, $\gamma_k$ and $t_k$ are random variables following a Gaussian distribution.

The ensemble average of a function $E(a)$, where $a$ is a random variable is defined as

\begin{equation} \label{undi6a}
	<E(a)> =  \int E(a)f(a)da.
\end{equation}

The function $f(a)$ is the probability distribution of $a$. The independence of the random variables allows to write the ensemble average $<E(\boldsymbol{r},\omega)>$ as the convolution of the single electron radiation with the probability density distributions of the electron beam, over offsets $f_{l}$, deflection angles $f_{\eta}$, the electron beam energy $f_{\gamma}$ and the longitudinal bunch profile $f_{t}$

\begin{equation} \label{undi45}
	<E(\boldsymbol{r},\omega)> = \int E_{\perp}(\boldsymbol{r},\omega,\boldsymbol{\eta},\mathbf{l},\gamma)e^{i\omega t}f_{\eta}(\boldsymbol{\eta})f_{l}(\mathbf{l})f_{\gamma}(\gamma)f_{t}(t)d\boldsymbol{\eta}d \mathbf{l} d \gamma dt.
\end{equation}

As a result, undulator radiation can be described mathematically in terms of statistical optics by averaging over the entire volume of the phase space of random parameters \cite{Trebushinin:22,Chubar22}. 

\subsection{Spatial coherence}\label{SRcoh}

The second-order coherence theory is the fundamental theory of optical coherence. It describes the correlation of electric field amplitudes \cite{born1999principles,mandel1995,goodman1988}. The second-order correlation is described by the mutual coherence function (MCF), defined as \cite{mandel1995}

\begin{equation} \label{coh1}
	\Gamma_t (\mathbf{r_1},\mathbf{r_2},t_1,t_2)=<E^*(\mathbf{r_1},t_1)E(\mathbf{r_2},t_2)>_t.	
\end{equation}

The MCF describes correlations between two electric field values $E(\mathbf{r_1},t_1)$ and $E(\mathbf{r_2},t_2)$ at different points in space $\mathbf{r_1}$ and $\mathbf{r_2}$ and times $t_1$ and $t_2$. The brackets $<...>$ denote the ensemble average. If the statistical process is stationary, quasi-stationary or ergodic, then the MCF can be written as a function that depends only on the time difference $\tau$ = $t_2-t_1$:
 
\begin{equation} \label{coh1a}
	\Gamma_t (\mathbf{r_1},\mathbf{r_2},\tau)=<E^*(\mathbf{r_1},t)E(\mathbf{r_2},t+\tau)>.
\end{equation}

In our $\{\mathbf{r}$-$\omega\}$-domain under consideration, the second order correlation function $\Gamma_{\omega} (\mathbf{r_1},\mathbf{r_2},\omega_1,\omega_2)$ can be written as follows 
\begin{equation} \label{coh2aa}
	\Gamma_{\omega} (\mathbf{r_1},\mathbf{r_2},\omega_1,\omega_2)=<E^*(\mathbf{r_1},\omega_1)E(\mathbf{r_2},\omega_2)>,	
\end{equation}
where functions $\Gamma_t (\mathbf{r_1},\mathbf{r_2},t_1,t_2)$ and $\Gamma_{\omega} (\mathbf{r_1},\mathbf{r_2},\omega_1,\omega_2)$ form a Fourier pair
\begin{equation} \label{coh11aa}
	\Gamma_t(\mathbf{r_1},\mathbf{r_2}, t_1, t_2) =\frac{1}{4\pi^2}\int_{-\infty}^{\infty} \Gamma_{\omega}(\mathbf{r_1},\mathbf{r_2},\omega_1,\omega_2)e^{i\omega_1 t_1}e^{-i\omega_2 t_2}d\omega_1d\omega_2.
\end{equation}
Taking into account the properties of synchrotron radiation described above, the second-order correlation function in the $\{\mathbf{r}$-$\omega\}$-domain can be split in the product of a spectral and a spatial factors \cite{geloni2008transverse}
\begin{equation} \label{coh11aaa}
	\Gamma_{\omega}(\mathbf{r_1},\mathbf{r_2},\omega_1,\omega_2)=G_{\omega}(\omega_1-\omega_2)G_{\perp}(\mathbf{r_1},\mathbf{r_2}),
\end{equation}
where $G_{\omega}(\omega_1-\omega_2)$ is the spectral correlation function which can be approximated by Dirac $\delta$-function $\delta(\omega_1-\omega_2)$ \cite{mandel1995}, and $G_{\perp}(\mathbf{r_1},\mathbf{r_2},\omega_1)$ is the cross-spectral density function (CSD)\cite{mandel1995}, which describes spatial correlations. From now on we will be concerned with the calculation of the cross-spectral density $G_{\perp}(\mathbf{r_1},\mathbf{r_2})$, where the frequency argument of the function is omitted and the quasi-monochromatic approximation is considered.
Using the mathematical description of the undulator source presented in section \ref{sourceM}, the CSD function (see supplementary material) is given by 
\begin{equation} \label{coh2a}
	\begin{aligned}
		G_{\perp}(\mathbf{r_1},\mathbf{r_2})=
		\frac{1}{2\pi}\int_{-\infty}^{\infty}d\Delta\omega~ E^*(\mathbf{r_1},\Delta\omega)E(\mathbf{r_2},\Delta\omega),
	\end{aligned}
\end{equation}

The fact that each monochromator has an intrinsically limited resolving power was taken into account in the derivation of Eq. \eqref{coh2a}\cite{geloni2008transverse}.
It is necessary to introduce a quantity for $G_{\perp}(\mathbf{r_1},\mathbf{r_2})$ that represents its efficiency. This quantity is the degree of transverse coherence (DoTC) $\zeta$, which characterises the spatial coherence of synchrotron radiation by a single number \cite{geloni2008transverse, mandel1995} and is described by

\begin{eqnarray}
	\zeta = \frac{\int_{-\infty}^{\infty}\int_{-\infty}^{\infty} d\mathbf{r_1} d\mathbf{r_2} \left|G_{\perp} (\mathbf{r_1},\mathbf{r_2})\right|^2}{\left|\int_{-\infty}^{\infty} d\mathbf{r} ~G_{\perp}(\mathbf{r},\mathbf{r}) \right|^2} ~.
	\label{coh14}
\end{eqnarray}

\section{Grating dispersion and spatial coherence degradation} 

The spatial coherence properties of the photon beam after interaction with a grating are described in detail in this section. The assumptions and simplifications of the previous sections are adopted. The most important simplification is the factorisation of the spatial and spectral components of the incident field amplitude. In addition, a fully coherent incident photon beam with a Gaussian spatial distribution is assumed. The influence of grating dispersion on spatial coherence is studied both for free space propagation and for focusing. The influence of aberrations in the focusing of the photon beam together with the grating dispersion is analysed in the following part. The variation of the spatial coherence properties of the photon beam as a function of the energy resolution of the grating monochromator is analysed in the last section.

\subsection{\label{sec:MOIg} Free-space propagation}
It has been shown that undulator radiation follows the same Gaussian random statistics as thermal light \cite{geloni2008transverse}. However, unlike thermal sources, which are completely incoherent, undulator sources are partially coherent, with a coherent spot size equal to the single-electron diffraction size.
From the sections \ref{sourceM} (Eq. \eqref{undi45}) and \ref{SRcoh} (Eq. \eqref{coh1}) it can be seen that the partial coherence of the undulator source is determined by the randomly distributed parameters $\boldsymbol{\eta}$, $\mathbf{l}$, $\gamma$ of the electron beam.   

The analysis presented here is limited to the effects associated with the grating. Effects associated with the random spatial and angular distribution of the electron beam are not considered, which assumes complete spatial coherence of the photon beam incident on the grating. As such, the photon beam incident on the grating with a given carrier frequency $\omega_0$ can be written as
  
\begin{equation} \label{deto1}
	E_{i}(\mathbf{r},t)=E_i(t)E_i(\mathbf{r})e^{-i\omega_0t},
\end{equation}

where we assumed that the temporal $E(t)$ and the spatial $E(\mathbf{r})$ components of the incident field amplitude $E_{i}(\mathbf{r},t)$ can be separated. 

\begin{figure}[tb]
	\centering
	\includegraphics[width=1.0\textwidth]{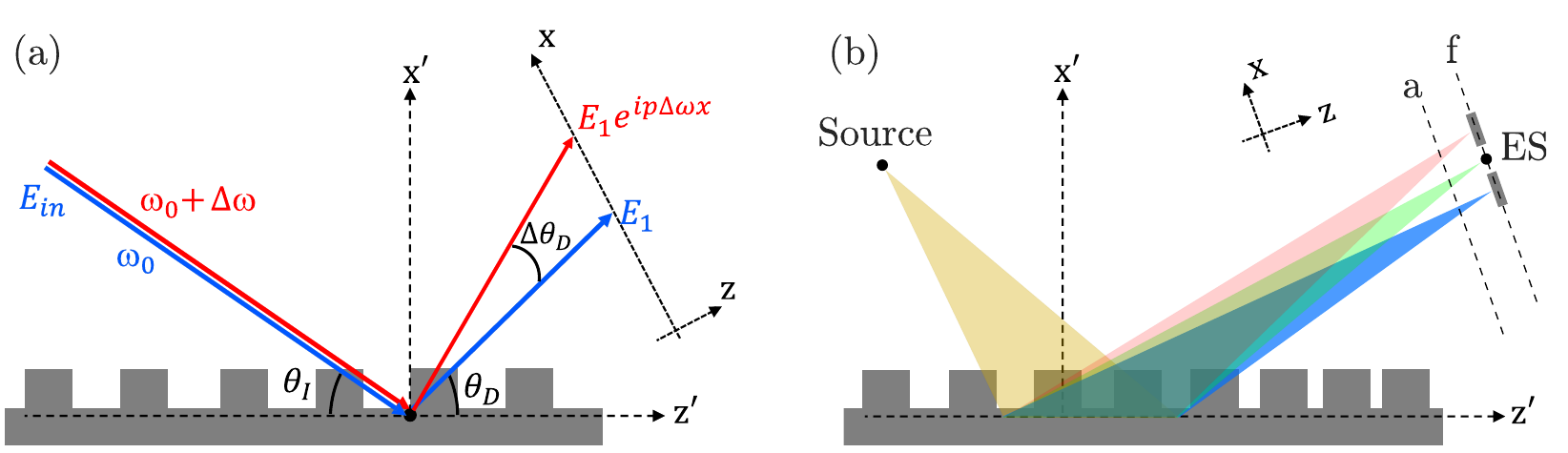}
	\caption{(a) Schematic of a plane grating. A photon beam with a frequency offset $\Delta\omega$ (red) from the carrier frequency $\omega_0$ (blue) exhibits an additional phase term, which depends on the specifications of the grating. (b) Schematic of a VLS grating. The polychromatic photon beam from the source is spectrally separated and focused by the VLS grating to the focal plane f at the ES position. To calculate defocus aberrations, the ES is placed out of focus at position a. Please note that all distances and angles in (a) and (b) are exaggerated. The spatial characteristics are considered in the $x$-plane, which is perpendicular to the propagation.} \label{FigG1a}
\end{figure}

In the following, only the one-dimensional spatial case is considered, in which the incident photon beam has the geometry shown in Fig. \ref{FigG1a}(a).

\begin{equation} \label{deto2}
	E_{i}(x,t)=E_i(t)E_i(x)e^{-i\omega_0t}.
\end{equation}

The electric field amplitude in the corresponding $\{x,\omega\}$ domain according to Eq. \eqref{inter1b} has the form 

\begin{equation}\label{deto3}
	E_i(x,\Delta\omega)=\int_{-\infty}^{\infty}dt ~E_i(t)E_i(x)e^{-i\omega_0t}e^{i \omega t} =E_i(\Delta\omega)E_i(x),
\end{equation}

where $\Delta\omega=\omega-\omega_0$. 



An incident electric field with Gaussian spatial and spectral distributions, characterised by their rms widths $\sigma_{x_0}$ and $\sigma_{\omega}$ respectively, has been considered for the following calculations. The Gaussian distributions are described by

\begin{equation}\label{moig1}
	E_i(x)=exp\bigg[-\frac{x^2}{2\sigma_{x_0}^2}\bigg],
\end{equation}
\begin{equation}\label{moig2}
	|E_i(\Delta\omega)|^2=exp\bigg[-\frac{\Delta\omega^2}{2\sigma_{\omega}^2}\bigg].
\end{equation}
Note that in Eq. \eqref{moig1} the amplitude of the incident electric field and in Eq. \eqref{moig2} the intensity of the field spectrum are considered. 

In the following, the Gaussian distributions described in Eqs. \eqref{moig1} and \eqref{moig2} are used to simplify the functional form of the CSD function. 
In this case, $E_i(\Delta\omega)$ can be considered as the ensemble average of the spectrum. The use of simple amplitude forms is sufficient for a quantitative analysis of the problem. According to Eq. \ref{coh2a}, upon using the following definitions 

\begin{eqnarray}
	&& x_1 = \bar{x} + \frac{\Delta x}{2} \cr
	&& x_2 = \bar{x} - \frac{\Delta x}{2} ~.
	\label{moi2aa}
\end{eqnarray}

the CSD of the electric field between points $x_1$ and $x_2$ can be defined in full generality (see Supplementary materials) by

\begin{equation}
	G_{\perp}(\bar{x},\Delta x) = \frac{1}{2\pi}   \int_{-\infty}^{\infty} d \Delta\omega ~ {E}^*\left(\bar{x} + \frac{\Delta x}{2},\Delta\omega\right){E}\left(\bar{x} - \frac{\Delta x}{2},\Delta\omega\right)~.
	\label{moi3}
\end{equation}

The DoTC is accordingly described by

\begin{eqnarray}
	\zeta = \frac{\int_{-\infty}^{\infty}\int_{-\infty}^{\infty} d\bar{x}~ d\Delta x~ \left|G_{\perp}(\bar{x},\Delta x)\right|^2}{\left|\int_{-\infty}^{\infty} d\bar{x} ~G_{\perp}(\bar{x},\Delta x=0) \right|^2} ~.
	\label{moi3a}
\end{eqnarray}

As already mentioned, we assume full spatial coherence of the beam incident on the grating, which can be seen by the substitution of $E_i(x, \Delta\omega)=E_i(x)E_i(\Delta\omega)$ in Eq. \eqref{moi3} and eq. \eqref{moi3a} 

\begin{equation}
	G_{\perp}(\bar{x},\Delta x) = \frac{1}{2\pi}   \int_{-\infty}^{\infty} d \Delta\omega ~ exp\bigg[-\frac{(\bar{x} + \frac{\Delta x}{2})^2}{2\sigma_{x_0}^2}\bigg]exp\bigg[-\frac{(\bar{x} - \frac{\Delta x}{2})^2}{2\sigma_{x_0}^2}\bigg]exp\bigg[-\frac{\Delta\omega^2}{2\sigma_{\omega}^2}\bigg]~,
	\label{moi3n}
\end{equation}

\begin{eqnarray}
	\zeta = \frac{\int_{-\infty}^{\infty}\int_{-\infty}^{\infty} dx_1 dx_2 \left|G_{i\perp} (\bar{x},\Delta x)\right|^2}{\left|\int_{-\infty}^{\infty} dx ~G_{i\perp}(\bar{x},\Delta x=0) \right|^2}=1 ~.
	\label{moig2a}
\end{eqnarray}

The CSD in Eq. \eqref{moi3n} is obtained by integrating over all individual frequencies within the incident energy bandwidth. It shows that the photon beam incident on the grating, which is characterised by Gaussian spatial and spectral distributions, has a degree of transverse coherence $\zeta = 1$ and is therefore fully spatially coherent. 

\begin{figure}[tb]
	\centering
	\includegraphics[width=1.0\textwidth]{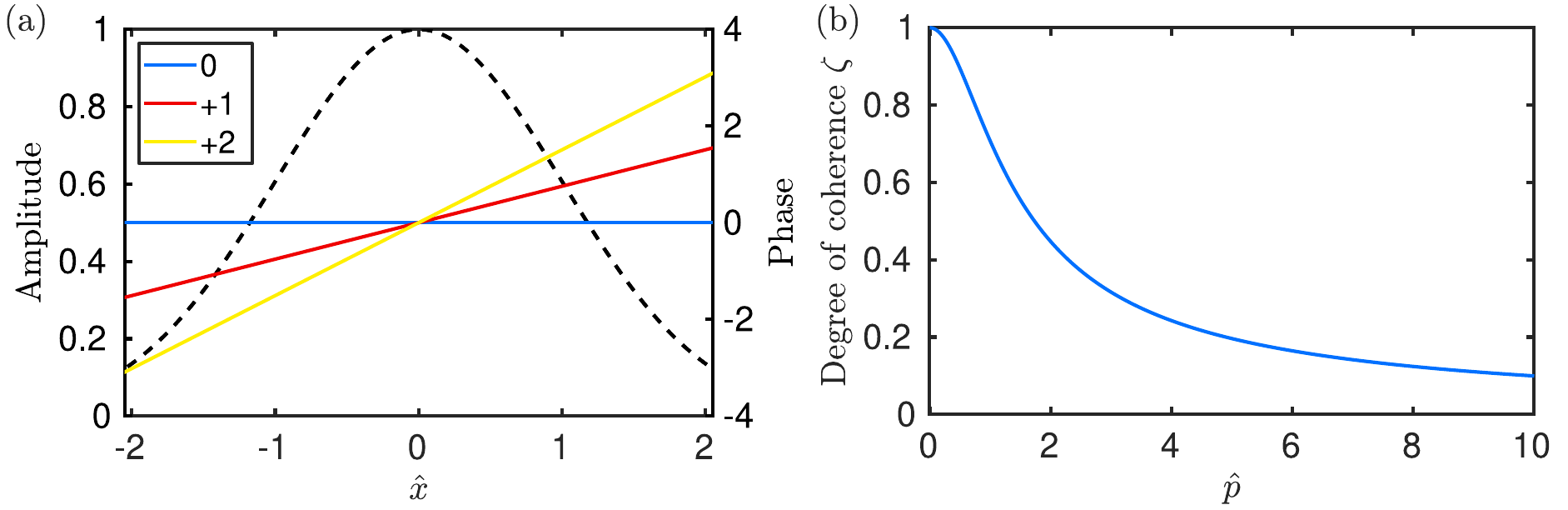}
	\caption{(a) Amplitude and phase distributions of the spatial part of the photon beam directly after the grating for three different frequencies (energies). The blue line corresponds to the phase distribution of the central frequency $\omega_0$ (e.\,g. 1200 eV), the red line for the frequency offset $\Delta\omega/\omega_0$ = +1$\cdot 10^{-5}$ (1200.01 eV) and the yellow line for the frequency offset $\Delta\omega/\omega_0$ = +2$\cdot 10^{-5}$ (1200.02 eV). The black dashed line corresponds to the amplitude distribution of the photon beam including all three frequencies. The horizontal $\hat{x}$-axis is normalized to the size of the beam $\sigma_x$. (b) DoTC of the photon beam direcly after the grating as a function of the normalised dispersion parameter $\hat{p}$ (see Eq. \eqref{moig5}).} \label{param0}
\end{figure}

An electric field incident on the grating with an energy offset $\hbar\Delta\omega$ from the resonant energy $\hbar\omega_0$ results in an angular increment, as shown in Fig. \ref{FigG1a}(a). 
As a result, it acquires an additional phase term in the $\{x,\Delta\omega\}$ domain (see Supplementary materials),

\begin{equation}\label{deto5}
	E_g(x,\Delta\omega)=E(x,\Delta\omega)e^{ip\Delta\omega x},
\end{equation}

where $p$=$\Delta k_x/\Delta \omega$ is the dispersion parameter of the grating.
The photon beam directly after the grating according to Eq. \ref{deto5} is described by

\begin{equation}\label{moig3}
	E_g(x,\Delta\omega)=exp\bigg[-\frac{x^2}{2\sigma_x^2}\bigg]E_i(\Delta\omega)e^{ip\Delta\omega x},
\end{equation}

where the photon beam width $\sigma_x=\frac{\theta_{D}}{\theta_{I}}\sigma_{x_0}$ is corrected for the "astigmatism" factor due to the difference in exit and entrance angles. 
In Eq. \eqref{moig3}, the diffracted electric field of each individual frequency $\Delta\omega$ contains a phase term with a certain tilt proportional to $\Delta\omega$ and $p$, with the exception of the carrier (resonant) frequency $\omega_0$. The tilt implies that the spatial and frequency components of the electric field are coupled. The amplitude and phase distribution of the spatial part of the electric field directly after the grating is shown in Fig. \ref{param0}(a) for three different energies. It shows that the photon beams strongly overlap spatially. However, the phase profiles of the individual photon beams are tilted with respect to the carrier (resonant) frequency ($\omega_0$).

Substitution of  Eq. \eqref{moig3} into Eq. \eqref{moi3} gives the expression for the CSD of the photon beam after the grating

\begin{equation}
	G_{g\perp}(\bar{x},\Delta x) = \frac{1}{\sqrt{2\pi}} \sigma_\omega \exp\left[-\frac{\sigma_\omega^2 p^2 (\Delta x)^2}{2 } \right] \exp\left[-\frac{\bar{x}^2 +(\Delta x)^2/4}{\sigma_x^2}\right] ~.
	\label{moig4}
\end{equation}

Further Substitution of the expression (Eq. \eqref{moig4}) into Eq. \eqref{moi3a} gives the DoTC 

\begin{equation}
	\zeta_g = \frac{1}{\sqrt{1+2p^2\sigma_{\omega}^2\sigma_{x}^2}} = \frac{1}{\sqrt{1+\hat{p}^2}},
	\label{moig5}
\end{equation}

where $\hat{p} = \sqrt{2} \sigma_\omega \sigma_x p$ is the normalised dispersion parameter. Equation \eqref{moig5} shows that the degree of transverse coherence is unity (full spatial coherence) for a monochromatic beam $\sigma_\omega$ = 0. The degree of transverse coherence as a function of the normalised dispersion parameter is shown in Fig. \ref{param0}(b). It can be seen that with increasing $\hat{p}$, the spatial coherence of the photon beam decreases significantly. The increase of $\hat{p}$ can be caused either by an increase of the incident energy bandwidth $\hbar\sigma_\omega$, the photon beam footprint on the grating $\sigma_x$/$\theta_{D}$, or the dispersion parameter $p$. 


Replacing the dispersion parameter $p$ by the grating parameters (see Supplementary materials), the degree of transverse coherence of the photon beam directly after the grating (assuming diffraction in the first order) is described by

\begin{equation}
	\zeta_g = \frac{1}{\sqrt{1+\frac{2\lambda^2\sigma_{\omega}^2\sigma_{x}^2}{{d}^2c^2\theta_{D}^2}}},
	\label{moig6}
\end{equation}

where $d$ is the groove spacing of the grating, $c$ is the speed of light and $\lambda$ is the wavelength corresponding to the carrier (resonant) frequency $\omega_0$.

In the framework of Gaussian-Schell Model (GSM) \cite{mandel1995,vartanyants2010} the CSD function $G_{GSM}$ is defined as follows
\begin{equation}
	G_{GSM}(\bar{x},\Delta x) = \sqrt{G_{\perp}(\bar{x}+\Delta x/2,0)}\sqrt{G_{\perp}(\bar{x}-\Delta x/2,0)}exp\bigg(-\frac{(\Delta x)^2}{L_g^2}\bigg).
	\label{moig5aa}
\end{equation}
By comparing Eq. \eqref{moig4} and Eq. \eqref{moig5aa} one can estimate the spatial coherence length of the diffracted field immidiately after the grating in the framework of GSM
\begin{equation}
	L_g = \frac{1}{\sigma_{\omega}p}.
	\label{moig5a}
\end{equation}
In Eq. \eqref{moig5a} it can be seen that the spatial coherence length depends only on the incoming bandwidth $\sigma_{\omega}$ and the grating parameter $p$.

In the following, the propagation of the diffracted electric field after the grating and its spatial coherence properties will be discussed. In this case, a slowly varying envelope of the field with respect to the frequency is assumed, as described by $E(x,z,\Delta\omega) = E_g(x,z_0,\Delta\omega) \exp(-i \omega z/c)$. The propagation of the diffracted field $E_g(x',\Delta\omega)$ in free-space up to a distance $z$ is given by the propagator 	
\begin{equation}\label{free0}
		P_z(x-x')= \sqrt{\frac{i\omega_0}{2\pi cz}}exp\bigg[\frac{i\omega_0(x-x')^2}{2cz}\bigg],
\end{equation}
and is defined as
\begin{equation}\label{free1}
	\begin{aligned}
		E_z(x,\Delta\omega)&= \int_{-\infty}^{\infty} E_g(x',\Delta\omega)P_z(x-x') dx'\\
		&=E_0exp\bigg[\frac{ix^2/\sigma_x-p^2\Delta\omega^2\sigma_x\hat{z}+2xp\Delta\omega\sigma_x}{2(\hat{z}-i\sigma_x)}\bigg]
	\end{aligned}
\end{equation}	
where $E_0$ contains all non-essential pre-integral factors, and $\hat{z}$ is defined as
\begin{equation}
	\hat{z}=\frac{cz}{\omega_0\sigma_{x}}.
\end{equation}

In Eq. \eqref{free1} a narrow bandwidth $\Delta\omega$ around $\omega_0$ is assumed. This assumption is valid until $\Delta \omega x^2/(cz) \ll 1$, otherwise we cannot consider the frequency $\omega$ = $\omega_0$ fixed anymore. Substitution of the propagator (Eq. \eqref{free1}) into Eq. \eqref{moi3} gives CSD function

\begin{equation}
	\begin{aligned}
		G_{z\perp}(\bar{x},\Delta x) = G_{z_0}exp\bigg[\frac{i\hat{z}\bar{x}\Delta x/\sigma_x -\bar{x}^2/\widetilde{p}-(\Delta x/2)^2}{\hat{z}^2+\sigma_{x}^2/\widetilde{p}}\bigg],
		\label{free3}
	\end{aligned}
\end{equation}	

where the new parameter $\widetilde{p}$ is defined as

\begin{equation}
	\widetilde{p}=1+2p^2\sigma_x^2\sigma_{\omega}^2 = 1 + \hat{p}^2,
\end{equation}

Interestingly, substitution of CSD (Eq. \eqref{free3}) into Eq. \eqref{moi3a} gives exactly the same expression as in Eq. \eqref{moig5} for DoTC directly after the grating $\zeta_z=\zeta_g$. This means that the degree of transverse coherence of the photon beam is preserved as it propagates in free-space after diffraction by the grating. However the coherence length of the beam now depending on the distance $z$ from the grating and can be estimated from Eq. \eqref{free3} in the framework of Gaussian-Schell Model \cite{mandel1995,vartanyants2010}  
\begin{equation}
	L_z = L_g\sqrt{\hat{z}^2/\sigma_{x}^2+2\hat{z}^2/L_g^2+1}.
	\label{free3a}
\end{equation}
For $z\rightarrow 0$, the coherence length is equal to the coherence length immediately behind the grating $L_z = L_g$.
%


\subsection{The influence of focusing}

For efficient monochromisation, the photon beam diffracted by the grating is focused into the plane of the exit slit aperture. Focusing can be achieved by using an additional focusing element after a plane grating or by using a self-focusing grating such as a Variable Line Spacing (VLS) grating (as shown in Fig.\ref{FigG1a}b). The latter is used in the following to describe the spatial coherence properties of the photon beam upon focusing. 

To simplify the mathematical description of a VLS grating, it can be represented by a plane grating in conjunction with a 'virtual lens' \cite{serkez2013grating,vartanyants2021}. This is sufficient for a qualitative analysis of the spatial coherence properties of a VLS grating. The approach simplifies the expressions for the diffracted field $E_g(x',\Delta\omega)$ and the CSD $G_{g\perp}(\bar{x},\Delta x)$. The virtual lens is mathematically described by the following transmission function

\begin{equation}\label{vls4}
	T_f(x,\omega_0)= exp\bigg[\frac{-i\omega_0x^2}{2cf}\bigg],
\end{equation}

where $f$ is the focal length of the VLS grating.
Using a propagator (Eq. \eqref{free0}), an electric field $E_f(x,\Delta\omega)$ in the focal plane of the virtual lens is described by

\begin{equation}\label{vls5}
	\begin{aligned}
		E_f(x,\Delta\omega)= \sqrt{\frac{i\omega_0}{2\pi cf}} \int_{-\infty}^{\infty} E_g(x',\Delta\omega)exp\bigg[\frac{-i\omega_0x'^2}{2cf}\bigg]exp\bigg[\frac{i\omega_0(x-x')^2}{2cf}\bigg] dx'&\\
		=\sqrt{\frac{i\sigma_x}{\sigma_{f}}} exp\bigg[\frac{-\Delta\omega^2}{4\sigma_{\omega}^2}\bigg]exp\bigg[\frac{i\omega_0x^2}{2cf}\bigg]exp\bigg[\frac{-x^2}{2\sigma_f^2}\bigg]exp\bigg[\frac{-p^2\Delta\omega^2\sigma_{x}^2}{2}\bigg]exp\bigg[\frac{p\Delta\omega x\sigma_{x}}{\sigma_{f}}\bigg],
	\end{aligned}
\end{equation}

where the width of the focused monochromatic beam $\sigma_f$ is defined by
\begin{equation}
	\sigma_{f}= \frac{fc}{\omega_0\sigma_{x}}.
	\label{vls5a}
\end{equation}
Note that the expression in Eq. \eqref{vls5a} is only valid for 1:1 focusing.

\begin{figure}[tb]
	\centering
	\includegraphics[width=1.0\textwidth]{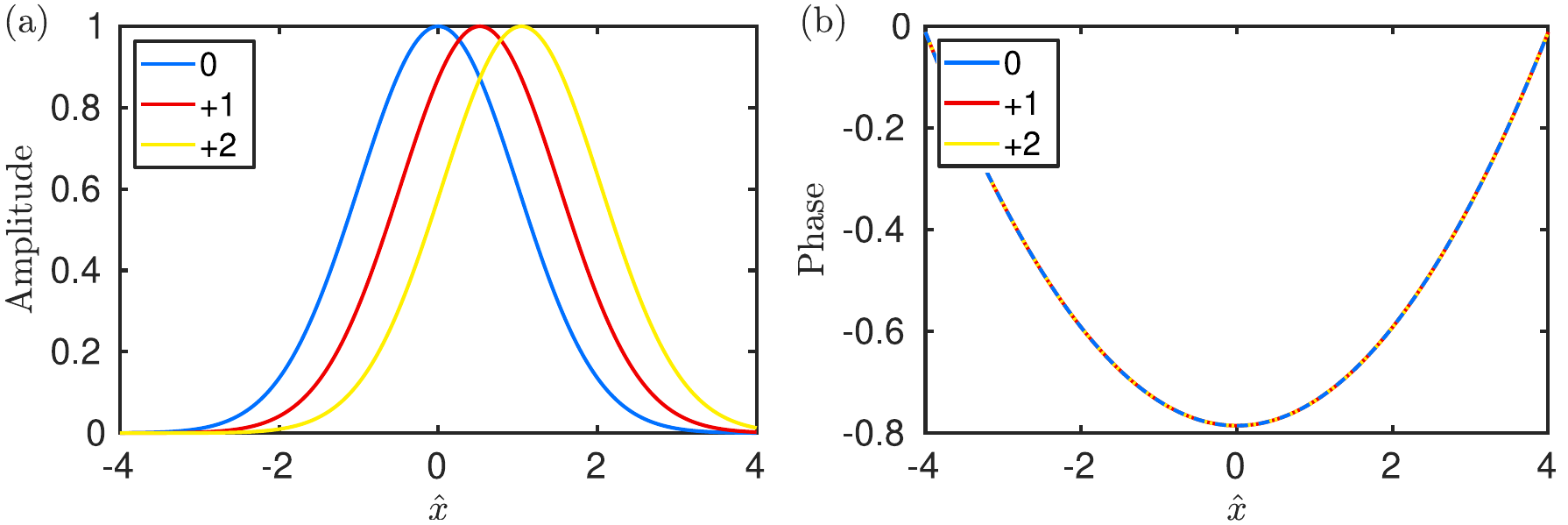}
	\caption{Amplitude and phase distributions of the spatial part of the photon beam at the exit slit plane for three different frequencies (energies). The blue line corresponds to the phase and amplitude distribution of the central frequency $\omega_0$ (e.\,g. 1200 eV), the red line for the frequency offset $\Delta\omega/\omega_0$ = +1$\cdot 10^{-5}$ (1200.01 eV) and the yellow line for the frequency offset $\Delta\omega/\omega_0$ = +2$\cdot 10^{-5}$ (1200.02 eV). The horizontal $\hat{x}$-axis is normalized to the size of the focused monochromatic beam $\sigma_f$. \label{Phase4}}
\end{figure}

The amplitude and phase distribution of the spatial part of the photon beam at the exit slit plane is shown in Fig. \ref{Phase4} for three different energies. In contrary to the case of free-space propagation (see Fig. \ref{param0}(a)), the phase distribution for the different energies is equal. However, the amplitude distribution of these photon beams are spectrally seperated. The degree of separation is determined by the dispersion parameter $p$. 
 
Substitution of Eq. \eqref{vls5} into Eq. \eqref{moi3} gives the expression for the CSD of the photon beam at the focal plane

\begin{equation}
	\begin{aligned}
		G_{f\perp}(\bar{x},\Delta x) = G_{f_0}exp\bigg[-\frac{i\bar{x}\Delta x}{\sigma_{f}\sigma_{x}}\bigg]exp\bigg(-\frac{[\bar{x}^2/\widetilde{p}+(\Delta x/2)^2]}{\sigma_{f}^2}\bigg).
		\label{vls6}
	\end{aligned}
\end{equation}
	
Suprisingly, further substitution of the expression (Eq. \eqref{vls6}) into Eq. \eqref{moi3a} gives the same expression as for the degree of transverse coherence directly after the grating (Eq. \eqref{moig5}) and during propagation $\zeta_f=\zeta_z=\zeta_g$ although the amplitude and phase distributions are different in both cases. This implies that the degree of spatial coherence of the photon beam does not change upon propagation or focusing and is hence preserved. Note that the result is the same if a plane grating is used in conjuction with a focusing mirror instead of a VLS grating. 
Similar to the previous calculations, the coherence length of the beam in the focus can be estimated from the CSD function in Eq. \eqref{vls6} 
\begin{equation}
	L_f = L_g\sqrt{\sigma_{f}^2/\sigma_{x}^2+2\sigma_{f}^2/L_g^2}.
	\label{vls6a}
\end{equation}


\subsection{Defocus aberration}

The following section describes the effect of defocus aberration on the spatial coherence properties of the photon beam after the grating. It is assumed that the defocus aberration has the largest effect on the beam properties. All other aberrations are neglected for simplicity. 

Defocus aberration is implemented by an additional phase factor

\begin{equation}\label{a2}
	P_a(x)=exp\bigg[\frac{iax^2}{\sigma_{x}^2}\bigg].
\end{equation}

Such aberrations can be considered in the out-of-focus plane, as shown in Fig. \ref{FigG1a}b. Substitution of Eq. \eqref{a2} into Eq. \eqref{vls5} gives the expression for the diffracted electric field with defocus aberration

\begin{equation}\label{a3}
	\begin{aligned}
		E^a_{f}(x,\Delta\omega)= \sqrt{\frac{i\omega_0}{2\pi cf}} \int_{-\infty}^{\infty} E_g(x',\Delta\omega)exp\bigg[\frac{-i\omega_0x'^2}{2cf}\bigg]P_a(x')exp\bigg[\frac{i\omega_0(x-x')^2}{2cf}\bigg] dx'&\\
		=E^a_{0} exp\bigg[\frac{-\Delta\omega^2}{4\sigma_{\omega}^2}\bigg]exp\bigg[\frac{i\omega_0x^2}{2cf}\bigg]exp\bigg[\frac{1}{(1-2ia)}\bigg(\frac{-x^2}{2\sigma_f^2}-\frac{p^2\Delta\omega^2\sigma_{x}^2}{2}+\frac{p\Delta\omega x\sigma_{x}}{\sigma_{f}}\bigg)\bigg].
	\end{aligned}
\end{equation}

The CSD function $G^a_{f\perp}(\bar{x},\Delta x)$ in the exit slit plane including defocus aberration is obtained by substitution of Eq. \eqref{a3} into Eq. \eqref{moi3}

\begin{equation}
	\begin{aligned}
		G^a_{f\perp}(\bar{x},\Delta x) = G_{a_0}exp\bigg[\frac{-i\bar{x}\Delta x}{\sigma_{f}\sigma_{x}}\bigg]exp\bigg(\frac{2ia\bar{x}\Delta x-\bar{x}^2 -(\Delta x/2)^2\widetilde{p}}{\sigma_{f}^2(4a^2+\widetilde{p})}\bigg).
		\label{a4}
	\end{aligned}
\end{equation}

For $a\rightarrow 0$, the CSD function $G^a_{f\perp}(\bar{x},\Delta x) = G_{f\perp}(\bar{x},\Delta x)$. The degree of transverse coherence $\zeta^a$ including defocus aberration is obtained by subtituting Eq. \eqref{a4} into Eq. \eqref{moi3a} which gives the same expression as in Eq. \eqref{moig5} ($\zeta^a$=$\zeta_g$), which is expected since DoTC does not change upon focusing or propagation. It confirms that the spatial coherence properties of the photon beam is preserved upon propagation and focusing which includes defocus aberration. 
Coherence length in the case of defocus aberration is obtained from the equation Eq. \eqref{a4} by the analogy with the previous calculations  
\begin{equation}
	L_a = L_g\sqrt{\sigma_{f}^2/\sigma_{x}^2+2\sigma_{f}^2/L_g^2+4a^2/\sigma_{x}^2}.
	\label{a4a}
\end{equation}
For $a\rightarrow 0$, the coherence length is equal to the coherence length in the focus $L_a = L_f$.
\subsection{Exit slit aperture and energy resolution}

The exit slit aperture is an integral part of a grating monochromator. It selects a certain spectral bandwidth of the incident photon beam and thus defines the resolving power of the monochromator. At the same time, it acts as a spatial filter and affects the spatial coherence properties of the monochromator. The effect of the exit slit aperture and its size on the spatial coherence properties of the photon beam is described below. Mathematically, the transmission function of the exit slit aperture can be approximated by a Gaussian function given by
 
\begin{equation}\label{es1}
	T_{s}(x)=exp(-\frac{x^2}{2\sigma_s^2}),
\end{equation}

where $\sigma_s$ is the root mean square of the slit amplitude function. Note that the exit slit size is the FWHM of $T_{s}(x)$. The intensity of the photon beam after the exit slit is defined as

\begin{equation}\label{es5}
	\begin{aligned}
		I_{es}(\Delta\omega)&= \bigg|\int_{-\infty}^{\infty} E_f(x,\Delta\omega)T_s(x) dx\bigg|^2\\
		&=\frac{2\pi\sigma_{f}\sigma_{x}}{\sigma_{s}\sqrt{\sigma_{f}^2/\sigma_{x}^2+(\sigma_{f}^2/\sigma_{s}^2+1)^2}}exp\bigg[-\frac{\Delta\omega^2}{2\sigma_{res}^2}\bigg],
	\end{aligned}
\end{equation}	

where $\sigma_{res}$ is the energy resolution of the monochromator that include incoming and transmitted bandwidth

\begin{equation}\label{es6}
	\begin{aligned}
		\sigma_{res}=\frac{1}{\sqrt{1/\sigma_{\omega}^2+1/\sigma_{\omega s}^2}}.	
	\end{aligned}
\end{equation}

In Eq. \eqref{es6}, the energy resoultion $\sigma_{\omega s}$ related to the exit slit aperture (transmitted bandwidth) is defined as

\begin{equation}\label{es7}
	\begin{aligned}
		\sigma_{\omega s}=\frac{1}{\sqrt{2\sigma_{x}^2p^2}}\bigg[\big(1+\frac{\sigma_{s}^2}{\sigma_{f}^2}\big)^2+\frac{\sigma_{s}^4}{\sigma_{f}^2\sigma_{x}^2}\bigg]^{\frac{1}{2}}\bigg[1+\frac{\sigma_{s}^2}{\sigma_{f}^2}\big(1+\frac{\sigma_{s}^2}{\sigma_{x}^2}\big)\bigg]^{-\frac{1}{2}}.	
	\end{aligned}
\end{equation}

If the exit slit aperture is fully closed $\sigma_{s}\rightarrow 0$, the normalised energy resolution is given by 

\begin{equation}\label{es7a}
	\begin{aligned}
		\frac{\sigma_{\omega s}}{\omega_0}=\frac{1}{\sqrt{2\sigma_{x}^2p^2\omega_0^2}}=\frac{1}{\sqrt{2\sigma_{x}^2\frac{4\pi^2}{\omega_0^2d^2\theta_{D}^2}\omega_0^2}}=\frac{1}{\pi\frac{2\sqrt{2}\sigma_{x}}{\theta_{D}d}}\approx\frac{1}{\pi\frac{\Delta x_f}{d}}=\frac{1}{\pi N_g},
	\end{aligned}
\end{equation}

where $\Delta x_f\approx 2\sqrt{2}\sigma_{x}/\theta_{D}$ (FWHM) is the footprint of the incident photon beam on the grating and $\Delta x_f/d=N_g$ is the number of illuminated grooves. In this case, the energy resolution of the monochromator is only determined by the total number of illuminated grooves. The maximum energy resolution $\sigma_{res}$ ($\sigma_{s}\rightarrow 0$) that can be obtained is described by  

\begin{equation}\label{es8}
	\begin{aligned}
		\sigma_{res}=\frac{\sigma_{\omega}}{\sqrt{1+2p^2\sigma_{x}^2\sigma_{\omega}^2}}=\frac{1}{\sqrt{1+\hat{p}^2}}=\frac{\sigma_{\omega}}{\sqrt{\widetilde{p}}}.
	\end{aligned}
\end{equation}

\begin{figure}[tb]
	\centering
	\includegraphics[width=1.0\textwidth]{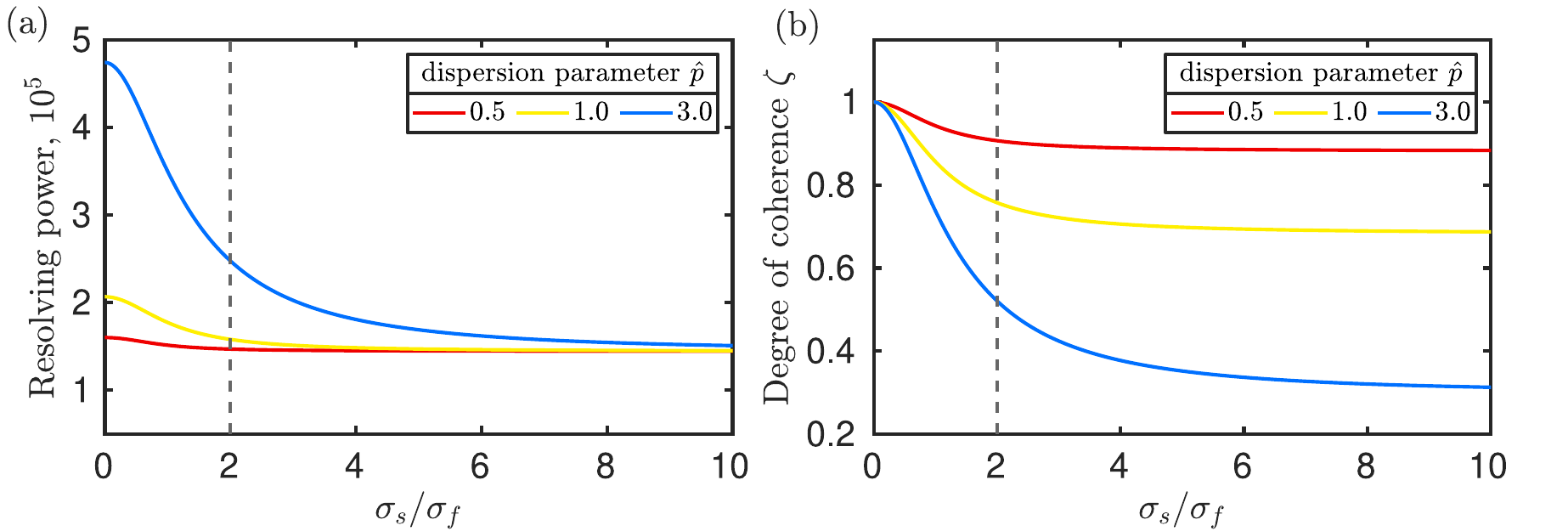}
	\caption{(a) Grating resolving power ($\omega_0/\sigma_{res}$, see Eq. \eqref{es6}) and (b) DoTC (see Eq. \eqref{es4}) as a function of the ratio of the exit slit size $\sigma_{s}$ and the size of the focused monochromatic beam $\sigma_{f}$ after the ES for different dispersion parameters $\hat{p}$.} \label{param3}
\end{figure}

The resolving power of the monochromator $\omega_0/\sigma_{res}$ as a function of $\sigma_{s}$ for different dispersion parameters $\hat{p}$ is shown in Fig. \ref{param3}(a). It can be seen that the resolving power increases as expected with decreasing $\sigma_{s}$ compared to $\sigma_{f}$. By increasing the dispersion parameter $\hat{p}$, the separation of the individual photon beams of different energies increases, resulting in an increasing resolving power for a given exit slit aperture size.

According to \cite{goodman1988}, the CSD function $G_{s\perp}(x_1,x_2)$ directly after the exit slit aperture is defined as

\begin{equation}\label{es2}
	G_{s\perp}(x_1,x_2)=T_{s}^*(x_1)T_{s}(x_2)G_{f\perp}(x_1,x_2),
\end{equation}

and in the {$\bar{x},\Delta x$}-domain by

\begin{equation}\label{es3}
	\begin{aligned}
		G_{s\perp}(\bar{x},\Delta x)=\frac{\sigma_{\omega}\sigma_{x}}{\sqrt{2\pi}\sigma_{f}}exp\bigg[\frac{-i\bar{x}\Delta x}{\sigma_{f}\sigma_{x}}\bigg]
		exp\bigg[\bar{x}^2\bigg(\frac{1}{\sigma_{f}^2\widetilde{p}}-\frac{1}{\sigma_s^2}\bigg)-\bigg(\frac{\Delta x}{2}\bigg)^2\bigg(\frac{1}{\sigma_{f}^2}+\frac{1}{\sigma_{s}^2}\bigg)\bigg].
	\end{aligned}
\end{equation}

The DoTC directly after the exit slit is given by

\begin{equation}
	\zeta_{es}= \sqrt{\frac{\sigma_s^2/(\sigma_f^2\widetilde{p})+1}{{\sigma_s^2/\sigma_f^2+1}}}.
	\label{es4}
\end{equation}

Figure \ref{param3}(b) shows the DoTC as a function of the rms exit slit aperture size $\sigma_{s}$ for different dispersion parameters $\hat{p}$. It can be seen that the degree of transverse coherence increases as expected with decreasing $\sigma_{s}$ compared to $\sigma_{f}$. By increasing $\hat{p}$ for a given $\sigma_{s}$, DoTC $\zeta_{es}$ decreases significantly. This is the opposite behaviour to the effect on the resolving power. As a result, the condition for the highest resolving power of the monochromator is accompanied by the lowest spatial coherence properties (for instance compare values for the slit openning of 2$\sigma_f$ for the same dispersion parameter $\hat{p}$, vertical dashed line Fig.\ref{param3}(a,b)). Note that the calculations assume 1:1 focusing without defocus aberration. It should also be noted that $\sigma_f$ is the effective size of the focused monochromatic photon beam, while the total photon beam size at the exit slit plane is proportional to its spectral bandwidth.

The effect of the exit slit aperture on the spatial coherence properties of the diffracted photon beam, including defocus aberration, is described below. Defocus aberration results in a translation of the focus along the optical axis out of the exit slit plane. Consequently, the size of the photon beam at the exit slit aperture is increased. The intensity distribution of the photon beam after the exit slit aperture is, in analogy to Eq. \eqref{es5}, given by

\begin{equation}\label{a6}
	\begin{aligned}
		I^a_{es}(\Delta\omega)&=  \bigg|\int_{-\infty}^{\infty} E^a_{f}(x,\Delta\omega)T_s(x) dx\bigg|^2\\
		&=\frac{2\pi\sigma_{f}\sigma_{s}^2}{\sqrt{(1+4a^2)(1/\sigma_{x}^2+\sigma_{f}^2/\sigma_s^2)+4(1-2a/\sigma_{x}\sigma_{s}+1/\sigma_{s}^2)}}exp\bigg[-\frac{\Delta\omega^2}{2\sigma_{res}^2}\bigg],&&
	\end{aligned}
\end{equation}	
 
and the bandpass $\sigma_{\omega s}$ related to the exit slit aperture size is defined as

\begin{equation}\label{a7}
	\begin{aligned}
		\sigma_{\omega s}&=\frac{1}{\sqrt{2\sigma_{x}^2p^2}}\bigg[1+\frac{\sigma_{s}^2}{\sigma_{f}^2}\big(1+\frac{\sigma_{s}^2}{\sigma_{x}^2}\big)\bigg]^{-\frac{1}{2}}\\
		&\times \bigg[\big(1+\frac{\sigma_{s}^2}{\sigma_{f}^2}\big)^2+\frac{\sigma_{s}^4}{\sigma_{f}^2\sigma_{x}^2}+4a\bigg(a+a\frac{\sigma_{s}^4}{\sigma_{x}^2\sigma_{f}^2}-\frac{\sigma_{s}^4}{\sigma_{f}^3\sigma_{x}}\bigg)\bigg]^{\frac{1}{2}}
		.	
	\end{aligned}
\end{equation}

Note, that the expression in Eq. \eqref{a7} takes the form of Eq. \eqref{es7} for ($a\rightarrow 0$).

\begin{figure}[tb]
	\centering
	\includegraphics[width=0.7\textwidth]{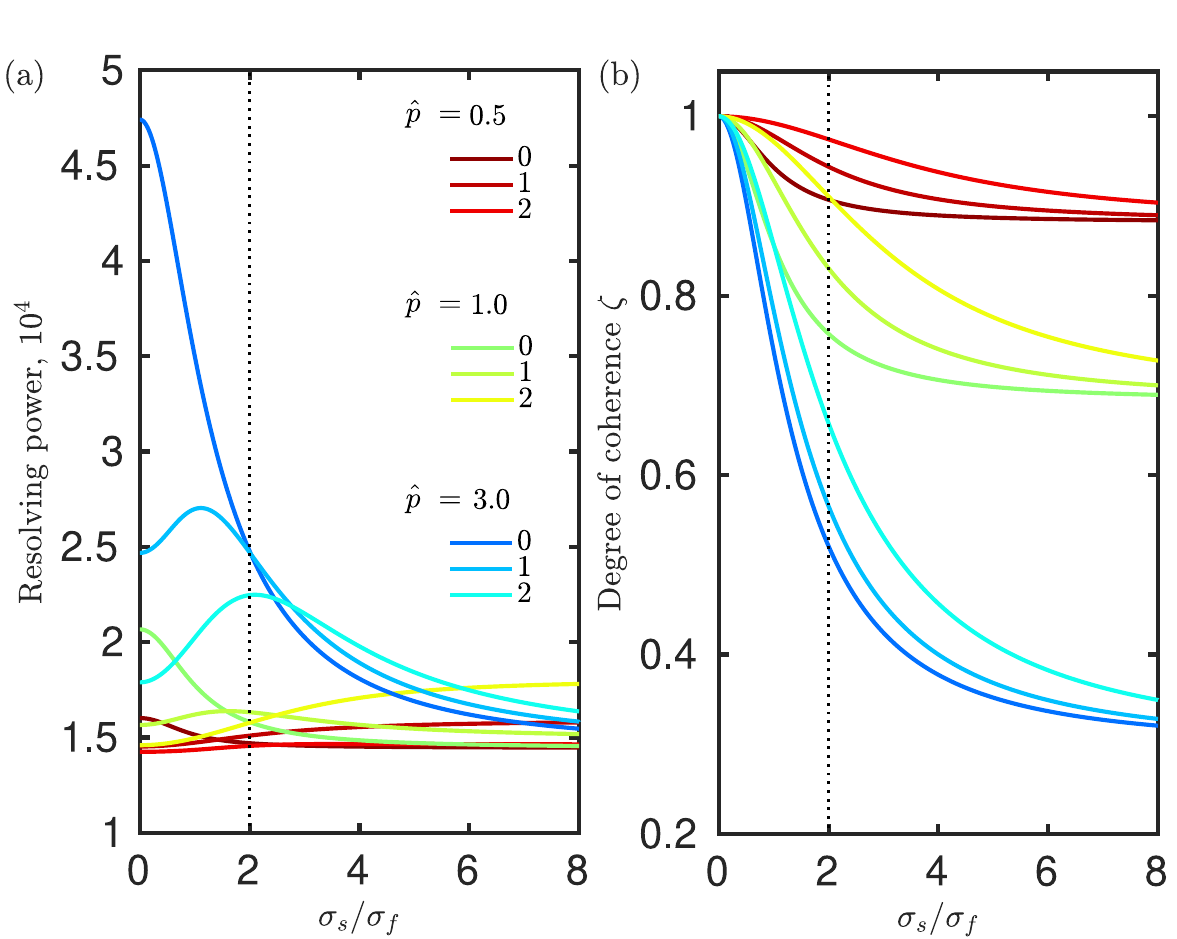}
	\caption{(a) Grating resolving power ($\omega_0/\sigma_{res}$, see Eq. \eqref{es6} and Eq. \eqref{a7}) as a function of the ratio of the exit slit size $\sigma_{s}$ and the size of the focused monochromatic beam $\sigma_{f}$ after the ES for different dispersion parameters $\hat{p}$ and defocus aberrations a. (b) DoTC as a function of the relation between the exit slit size $\sigma_{s}$ and the size of the focused monochromatic beam $\sigma_{f}$ after the ES for different dispersion parameters $\hat{p}$ and defocus aberrations a (see Eq. \eqref{a5}). The ES is placed out of focus for the calculations (see Fig. \ref{FigG1a}).} \label{aber2}
\end{figure}

The resolving power for different grating settings is shown in Fig. \ref{aber2}(a). It can be seen that, for a given exit slit aperture and dispersion parameter $\hat{p}$, the resolving power decreases significantly with increasing defocus aberration. The reason for this is the decreasing spatial separation of the photon beams of individual energies out of focus illuminating the exit slit aperture. For large exit slit sizes or small focus sizes, the effect of aberration is less pronounced.

The CSD function of the photon beam directly after the exit slit aperture is defined according to Eq. \eqref{es2} and the DoTC according to Eq. \eqref{moi3a} which is given by

\begin{equation}
	\zeta^a_{es}= \sqrt{\frac{\sigma_s^2/\sigma_f^2+\widetilde{p}+4a^2}{{\widetilde{p}(\sigma_s^2/\sigma_f^2+1)+4a^2}}}.
	\label{a5}
\end{equation}

The degree of transverse coherence for different grating settings is shown in \ref{aber2}(b). It can be seen that the spatial coherence increases with increasing defocus aberration for a given exit slit aperture size and dispersion parameter $\hat{p}$. The increased size of the photon beam at the plane of the exit slit aperture due to defocus aberration leads to a stronger clipping of the photon beam and thus to a higher degree of transverse coherence as expected. Consequently, the effect of defocus aberration favours the spatial coherence properties of the photon beam after the exit slit. At the same time, however, the resolving power is reduced. This is similar to the case without defocus aberration.

\section{Discussion}\label{sec:PhysExp}

\begin{figure}[tb]
	\centering
	\includegraphics[width=1.0\textwidth]{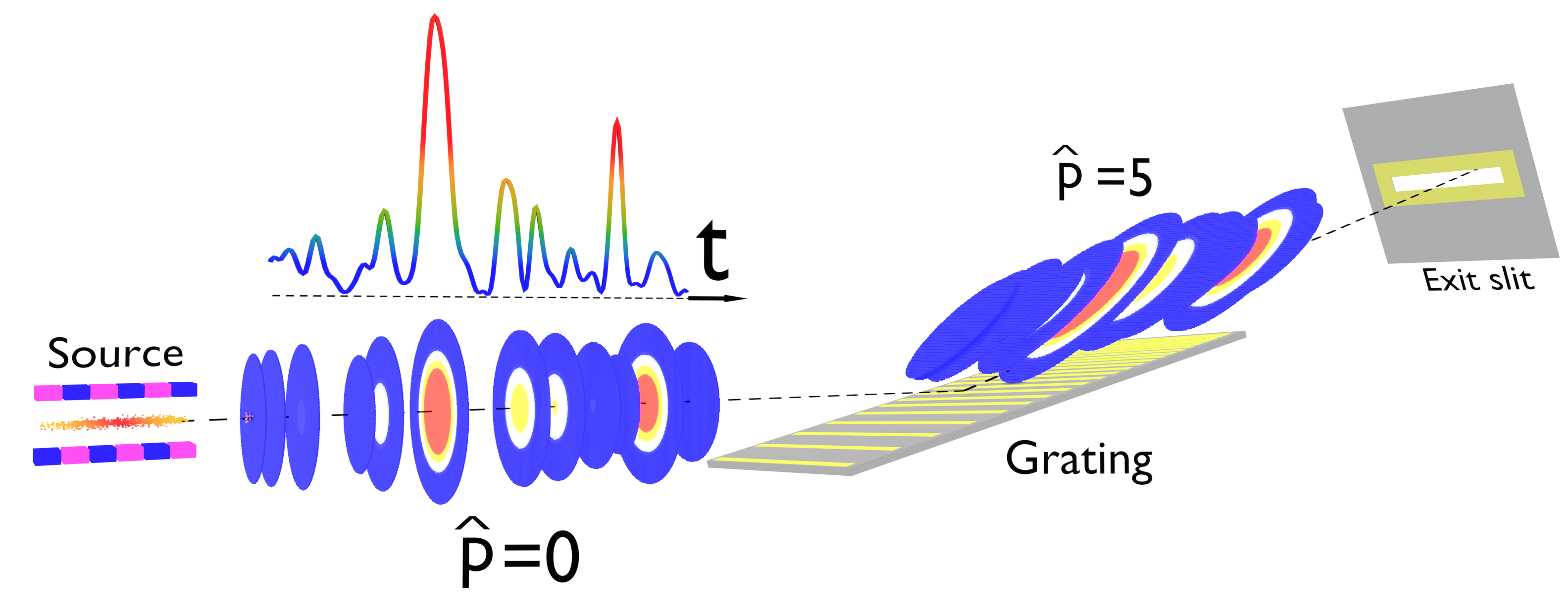}
	\caption{Visual representation of the SR pulse intensity in the $\{\mathbf{r},t\}$-domain before and after the interaction with a grating, where normalized dispersion parameter of the pulse with PFT $\hat{p}$ = 5. \label{PFT}}
\end{figure}
The mathematical description of the interaction of undulator radiation with a grating and the spatial coherence degradation caused by dispersion has been described in the previous sections in the $\{x,\omega\}$ domain. There is a one-to-one correspondence between the $\{x,\omega\}$ and the $\{x,t\}$ domains. This means that the effect of spatial coherence degradation can also be described in the $\{x,t\}$ domain. It has been shown that the diffracted field after the grating has an additional phase factor in the $\{x,\omega\}$ domain (Eqs. \eqref{moig3}, \eqref{deto5}). In the $\{x,t\}$ domain, the diffracted field contains an additional shift $t-px$.
This electric field is described by

\begin{equation}\label{deto6}
	\begin{aligned}
		E_g(x,t)=\frac{1}{2\pi}\int_{-\infty}^{\infty}E_i(\Delta\omega)E_i(x)e^{ip\Delta\omega x}e^{-i \omega t}d\Delta\omega\approx
		E_i(t-px)E_i(x)e^{-i\omega_0t}.
	\end{aligned}
\end{equation}

From Eq. \ref{deto6} it follows that the temporal field now also spatially dependent, which is known as pulse front tilt phenomenon (PFT)\cite{treacy1969,bor1993,hebling1996derivation,akt2005,weiner2011}, i.e. the space and time domains are coupled (space-frequency coupling in the $\{x,\omega\}$ domain). 

Prior to the interaction of the undulator radiation with the grating, we assumed only one spatial mode, which implies full spatial coherence of the photon beam (see Eq. \eqref{moig2a} and Fig.\ref{PFT}). However, due to the properties of synchrotron undulator radiation, it has a large number of longitudinal or spectral modes (Fig.\ref{PFT}), implying low temporal coherence. The number of these modes can be estimated from the intensity fluctuations of the undulator radiation (roughly equal to the number of spikes, see Fig. \ref{Fluc1}). The dispersion effect of the grating leads to a redistribution of the total number of modes, whereby temporal (or spectral) modes are partially converted into spatial modes, resulting in the phenomenon of PFT (Fig.\ref{PFT}). The degree of redistribution is determined by the dispersion parameter $p$. In this case the effect of the time delay of a pulse in the dispersion plane is observed for individual temporal modes of the undulator radiation, while each individual mode is spatially fully coherent. The result is multiple spatial modes observed in the dispersion plane for a given bandwidth, statistically causing the decoherence effect. 

The statistical analysis presented applies under the assumption that the grating monochromator is unable to resolve individual spectral spikes. If the grating has specifications high enough to resolve individual spectral spikes, the redistribution of modes after the grating would result in a complete conversion of spectral modes into spatial. In this case, only one longitudinal coherence mode would be present and the photon beam would be fully spatially incoherent. 

\section{Conclusion}\label{sec:Diss}

The presented analysis shows that a grating used for grating monochromators significantly affects the spatial coherence properties of the photon beam. This effect can be attributed to the properties of synchrotron radiation pulses in combination with the dispersion properties of the grating. It has been shown that the higher the dispersion parameter $\hat{p}$ of the grating, the lower the spatial coherence of the diffracted field. The dispersion parameter $\hat{p}$ depends on both the photon energy and the bandwidth of the incident radiation, as well as the footprint of the beam on the grating, its line density and the used diffraction order. With strong dispersion, a spatial coherence degradation of ~80$\%$ can be achieved. 

It has been shown that the reduced degree of spatial coherence of the diffracted field after the grating is maintained upon propagation and focusing. This is due to the fact that as the beam propagates, only the scales change, not the phase dependencies. In the focal plane, the phase tilt is cancelled for each individual frequency component, but the individual beams are strongly spatially separated, which strongly affects the spatial coherence. 

The effects of grating dispersion on the spatial coherence and resolution of the monochromator have been investigated in conjunction with the exit slit aperture. It is shown that the resolving power of the monochromator and the spatial coherence of the dispersed beam can be significantly increased by closing the exit slit. However, it has also been shown that for a given exit slit aperture opening, the degree of spatial coherence is inversely proportional to the resolving power of the monochromator. This means that after passing through the monochromator a highly dispersed beam will have the lowest spatial coherence.

Defocus aberration alone cannot affect the spatial coherence of the photon beam. However, in combination with the monochromator exit slit aperture, it can. The aberration defocuses the photon beam, resulting in over-illumination of the exit slit aperture for a given exit slit aperture size. This results in an increase in spatial coherence and a decrease in the resolving power of the grating. 

The presented results can help in mitigating the impact of spatial coherence degradation at synchrotron beamlines that employ grating monochromators. This can be achieved by selecting appropriate grating parameters and settings. 

It would be of great interest to investigate the degradation of spatial coherence caused by a grating for partially coherent X-ray beams at future 4GSRs, which are expected to provide high spatial coherence of the source. This case goes beyond the ideal model considered in this work and imposes complications on the functional forms of the spatial correlation functions. It is expected that the effect of spatial coherence degradation will become even more pronounced than in the ideal case of a fully spatially coherent source. The analysis of the spatial coherence degradation of a grating illuminated by a partially spatially coherent beam will be described in a forthcoming paper.

\section{Supplementary materials}
\subsection{\label{sec:inrter} Kinematical scattering and dispersion }

\begin{figure}[tb]
	\centering
	\includegraphics[width=1.0\textwidth]{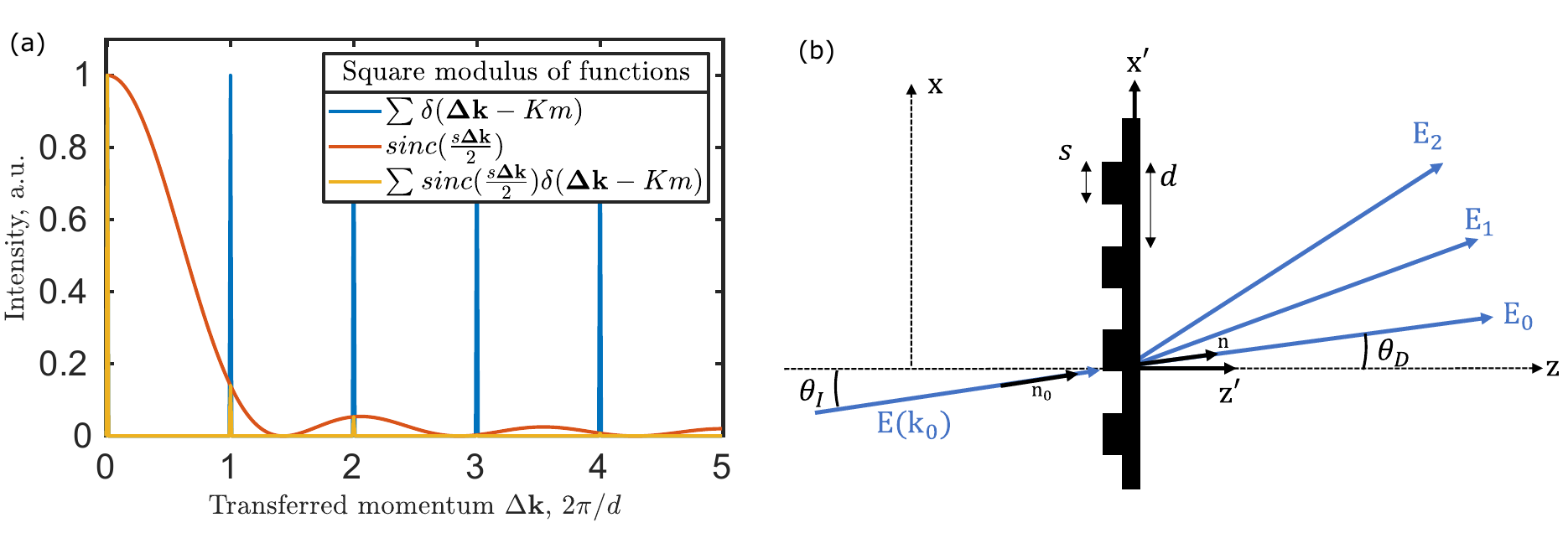}
	\caption{(a) A beam propagating through a grating with a period $d$ and size of the period $s$. (b) Various functions for the diffracted field.} \label{FigG0}
\end{figure}

In this section we would like to remind a general concept while considering a scattering problem from the diffractive element. In our case a diffractive element, in particular a grating is represented as the number of point scatters in the medium, placed in the periodic manner. As the first step towards the reperesentation of the problem a wave equation within the scalar theory is considered for the component of the electric filed $E(\mathbf{r},t)$ without presense of sources but for inhomogeneous media of non-magnetic materials
\begin{equation}\label{inter0}
	\big[\epsilon(\mathbf{r})\mu_0 \frac{\partial^2}{\partial t^2}-\nabla^2\big]E(\mathbf{r},t) = 0,
\end{equation}
where $\epsilon(\mathbf{r})$ is the electrical permittivity and $\mu_0$ is the magnetic permeability of the medium. In the following the complex scalar electromagnetic wave $E(\mathbf{r},t)$ is reperesnted as the continuos superposition of the monochromatic componets, via a Fourier integral
\begin{equation}\label{inter1}
	E(\mathbf{r},t) = \frac{1}{2\pi} \int_{-\infty}^{\infty} E(\mathbf{r},\omega)e^{-i \omega t} d\omega.
\end{equation}
Taking into account equations \eqref{inter0} and \eqref{inter1} the inhomogeneous (with respect to medium) Helmholtz equation, which describes the interaction of monochromatic X-rays with matter, is written in the following form
\begin{equation}\label{inter2}
	[\nabla^2 + k^2n^2(\mathbf{r},\omega)]E(\mathbf{r},\omega) = 0,
\end{equation}
where $k=\omega c$ is the wave vector and $n(\mathbf{r},\omega)$ is the frequency-dependent refractive index
\begin{equation}\label{inter3}
	n(\mathbf{r},\omega)=c\sqrt{\mu_0\epsilon(\mathbf{r},\omega)}.
\end{equation}
Considering a single point scatter in the presence of an incident monochromatic scalar electromagnetic filed one can define a Green function, which is the field that is scattered from such a point. For this equation \eqref{inter2} usually is rewritten in the following form
\begin{equation}\label{inter4}
	[\nabla^2 + k^2]E(\mathbf{r},\omega) = k^2[1 - n^2(\mathbf{r},\omega)]E(\mathbf{r},\omega),
\end{equation}
which means that in vacuum right-hand side of this equation vanishes, leaving the homogeneous Helmoltz equation for scalar fields. In the presence of the point scatter (Delta function located at the origin of coordinates) the following equation can be written
\begin{equation}\label{inter5}
	[\nabla^2 + k^2]G(\mathbf{r},\omega) = -\delta(\mathbf{r}),
\end{equation}
which solution is the free-space Green function represented as outgoing spherical wave from the point scatter
\begin{equation}\label{inter6}
	G(\mathbf{r},\omega) = \frac{1}{4\pi}\frac{e^{ik|\mathbf{r}|}}{|\mathbf{r}|}.
\end{equation}
With the help of Green function \eqref{inter6} and Eq. \eqref{inter4} the integral form of scattering equation can be written
\begin{equation}\label{inter7}
	E(\mathbf{r},\omega) = E_{in}(\mathbf{r},\omega) - \int G(\mathbf{r}-\mathbf{r}',\omega)k^2[1 - n^2(\mathbf{r}',\omega)]E(\mathbf{r}',\omega) d\mathbf{r}',
\end{equation}
where $E_{in}(\mathbf{r},\omega)$ is the incident unscattered field.
The integrand in equation \eqref{inter7} is non-zero only within the scattering volume where the refractive index is not unity. First Born approximation allows to assume that X-ray field inside the scattering volume is only slightly different from the field that would have been at each $\mathbf{r}'$ point in the volume in the absence of the scatter. The latter allows to rewrite the integral equation \eqref{inter7} in the following form
\begin{equation}\label{inter8}
	E(\mathbf{r},\omega) = E_{in}(\mathbf{r},\omega) - \int G(\mathbf{r}-\mathbf{r}',\omega)k^2[1 - n^2(\mathbf{r}',\omega)]E_{in}(\mathbf{r}',\omega) d\mathbf{r}',
\end{equation}
which is the expression for the total field as the a sum of non-scatered incident part $E_{in}(\mathbf{r},\omega)$ and the part which acounts the scattering by the medium. A single-scattering is assumed witihin this kinematical theory where the incident wave-field is scattered only once by a single point within the media. 
Now for the simplicity we will write Eq. \eqref{inter8} assuming the incident wave-filed in the form of a monochromatic plane wave $E_{in}(\mathbf{r},\omega) = E_{0}e^{i\mathbf{k_{0}}\mathbf{r}}$, and the Green fuction as in Eq. \eqref{inter6}, obtaining
\begin{equation}\label{inter9}
	E(\mathbf{r},\omega) = e^{i\mathbf{k_{0}}\mathbf{r}} - \frac{k^2}{4\pi}\int \frac{e^{ik|\mathbf{r}-\mathbf{r}'|}}{|\mathbf{r}-\mathbf{r}'|} [1 - n^2(\mathbf{r}',\omega)]e^{i\mathbf{k_{0}}\mathbf{r}'} d\mathbf{r}'.
\end{equation}
In the equation \ref{inter9}, nonessential pre-integral factors were omitted. We will also use far-filed approxiation where the observation point $\mathbf{r}$ is located at a distances much greater than the size of the scattering medium ( $\approx|\mathbf{r}'|$) so that $|\mathbf{r}|\gg |\mathbf{r}'|$. In this case the term in exponential function simplidfies as
$|\mathbf{r}-\mathbf{r}'|$=$\sqrt{(\mathbf{r}-\mathbf{r}')^2}$=$\sqrt{|\mathbf{r}|^2-2\mathbf{r}\mathbf{r}'+|\mathbf{r'}|^2}$$\approx$$|\mathbf{r}|\sqrt{1-2\mathbf{r}\mathbf{r}'/|\mathbf{r}|^2}$ and with the help of binominal approximation $\sqrt{1-2\mathbf{r}\mathbf{r}'/|\mathbf{r}|^2}$$\approx$$1-\mathbf{r}\mathbf{r}'/|\mathbf{r}|^2$ so that the Green function can be written as
\begin{equation}\label{inter10}
	G(\mathbf{r}-\mathbf{r}',\omega) = \frac{1}{4\pi}\frac{e^{ikr}}{r}e^{-\frac{ik\mathbf{r}\mathbf{r}'}{r}},
\end{equation}
where $|\mathbf{r}|$ was replaces by $r$.
In this way Eq. \eqref{inter9} simplifies to 
\begin{equation}\label{inter11}
	E(\mathbf{r},\omega) = e^{i\mathbf{k_{0}}\mathbf{r}} + \frac{k^2}{4\pi}\frac{e^{ikr}}{r}\int e^{-\frac{ik\mathbf{r}\mathbf{r}'}{r}} [n^2(\mathbf{r}',\omega) - 1]e^{i\mathbf{k_{0}}\mathbf{r}'} d\mathbf{r}'.
\end{equation}
Second part of this equation \eqref{inter11} represents a scttered amplitude, which can be rewritten as
\begin{equation}\label{inter12}
	E_{d}(\mathbf{\Delta k},\omega) =\frac{k^2}{4\pi}\frac{e^{ikr}}{r}\int [n^2(\mathbf{r}',\omega) - 1]e^{-i \mathbf{\Delta k}\mathbf{r}'} d\mathbf{r}'.
\end{equation} 
where $\mathbf{\Delta k}=k\mathbf{n}-\mathbf{k_{0}}$=$k(\mathbf{n}-\mathbf{n_{0}})$ is the vector difference between incident and scattred wave vectors (transferred momentum) and $\mathbf{n}=\mathbf{r}/|\mathbf{r}|$ is the unit vector, pointing to the observation direction. Generally speaking Eq. \eqref{inter12} represents a Fourier transform of the function $T(\mathbf{r}',\omega)=[n^2(\mathbf{r}',\omega) - 1]$ with the position dependent refractive index $n(\mathbf{r}',\omega)$ of the scattering medium. In the case of plane grating the function  $T(\mathbf{r}',\omega)$=$R(\mathbf{r}',\omega)$ is called a transmission or reflection function of the grating (depending on geometry)  and can be written in the same manner as for crystals (i.e. scattering from a periodic potential)
\begin{equation}\label{inter13}
	T(\mathbf{r}')=T_p(\mathbf{r}')T_s(\mathbf{r}'),
\end{equation}
where $T_{p}(\mathbf{r}')$ is periodic scattering potential (function of the grating grooves) and $T_{s}(\mathbf{r}')$ is the shape function of the crystal (shape of the grating). In the following a general approach will be given, such that vector $\mathbf{r}'$ is kept. When diffraction is considered on one- or two-dimensional structure, $\mathbf{r}'$ equals to $x$ or $(x,y)$ correspondingly. Since $T_{p}(\mathbf{r}')$ is periodic it can be described by the Fourier series
\begin{equation}\label{inter14}
	T_p(\mathbf{r}')= \sum_{m=-\infty}^{\infty} t_{m}e^{iKm\mathbf{r}'},
\end{equation} 
where $K=\frac{2\pi}{d}$ is the spatial wave number for the periodic function of the grating with spacing $d$ and Fourier coefficients $t_{m}$, 
\begin{equation}\label{inter15}
	t_{m}= \frac{1}{d}\int_{-d/2}^{d/2} T_p(\mathbf{r}')e^{-imK\mathbf{r}'}d\mathbf{r}'. 
\end{equation}
At first, for simplicity the periodic function can be choosen as periodic Dirac delta function or Dirac comb
\begin{equation}\label{inter14a}
	T_p(\mathbf{r}')= \Sha (\mathbf{r}')=\sum_{m=-\infty}^{\infty}\delta(\mathbf{r}'-mD)=\frac{1}{d}\sum_{m=-\infty}^{\infty} e^{iKm\mathbf{r}'},
\end{equation} 
which coefficient $t_{m}=1/d$. Assuming infinite size of the grating, such that a shape function equals to one everywhere $T_s(\mathbf{r}')$=$1$, Eq. \eqref{inter12} simplifes to
\begin{equation}\label{inter15a}
	E_{d}(\mathbf{\Delta k},\omega) =\frac{k^2}{4\pi}\frac{e^{ikr}}{r}\frac{1}{d} \int \sum_{m=-\infty}^{\infty} e^{iKm\mathbf{r}'} e^{-i \mathbf{\Delta k}\mathbf{r}'} d\mathbf{r}'.
\end{equation} 
By changing the order of summation and integration one gets Dirac comb in $\mathbf{\Delta k}$-space as a result
\begin{equation}\label{inter15b}
	\begin{aligned}
		E_{d}(\mathbf{\Delta k},\omega)& =\frac{k^2}{4\pi}\frac{e^{ikr}}{r} \frac{1}{d} \sum_{m=-\infty}^{\infty} \int e^{i(Km- \mathbf{\Delta k})\mathbf{r}'} d\mathbf{r}' \\ 
	&	=\frac{k^2}{4\pi}\frac{e^{ikr}}{r} \frac{1}{d}\sum_{m=-\infty}^{\infty} \delta(\mathbf{\Delta k}-Km),
	\end{aligned}
\end{equation}
Normalized intesity function $I_{d}(\mathbf{\Delta k},\omega)$=$|E_{d}(\mathbf{\Delta k},\omega)|^2$ in this case is shown in Fig. \ref{FigG0}(a)
Non-zero contribution for the diffracted wave implies the condition at which the wave vector difference $\mathbf{\Delta k}$ should match the integer number of grating periods
\begin{equation}\label{inter16}
	\mathbf{\Delta k}=\frac{2\pi}{d}m,
\end{equation}
which brings a general grating condition. If we choose a system such as shown in Fig. \ref{FigG0}(b) then we will see diffracted intensity peaks at the certain angles dictadet by the grating equation
\begin{equation}\label{inter17}
	k[sin(\theta_{D})-sin(\theta_I)]=\frac{2\pi}{d}m, 
\end{equation}
where difraction order $m=0, \pm 1, \pm 2 ...$ or
\begin{equation}\label{inter17a}
	sin(\theta_{D})-sin(\theta_I)=\frac{\lambda}{d}m. 
\end{equation}
According to Eq. \eqref{inter14}, taking a rectangular function with size $s$ as a shape function of one period as shown in Fig.,
\begin{equation}\label{inter17b}
	T_p(\mathbf{r}')=rect\bigg(\frac{\mathbf{r}'}{s}\bigg)=\begin{cases}
		0,  & \mbox{if } |\mathbf{r}'|> s/2 \\
		1/2, & \mbox{if } |\mathbf{r}'|= s/2 \\
		1, & \mbox{if } |\mathbf{r}'|< s/2 \\
	\end{cases}
\end{equation} 
results in $t_{m}$ coefficients in the form of a sinc function  
\begin{equation}\label{inter15aa}
	t_{m}= \frac{1}{d}\int_{-d/2}^{d/2} rect(\frac{\mathbf{r}'}{s})e^{-imK\mathbf{r}'}d\mathbf{r}'=\frac{1}{d}\frac{sin(sKm\frac{1}{2})}{Km\frac{1}{2}}=\frac{s}{d}sinc(\frac{sKm}{2}). 
\end{equation}
In this case, diffracted amplitude $E_{d}(\mathbf{\Delta k},\omega)$ represented as a Dirac comb function, which is modulated by the coefficient $t_{m}$ 
\begin{equation}\label{inter15bb}
	\begin{aligned}
		E_{d}(\mathbf{\Delta k},\omega)& =\frac{k^2}{4\pi}\frac{e^{ikr}}{r}\frac{1}{d} \sum_{m=-\infty}^{\infty} \frac{sin(sKm\frac{1}{2})}{Km\frac{1}{2}} \int e^{i(Km- \mathbf{\Delta k})\mathbf{r}'} d\mathbf{r}' \\ 
		&=\frac{k^2}{4\pi}\frac{e^{ikr}}{r}\frac{s}{d} \sum_{m=-\infty}^{\infty} \frac{sin(sKm\frac{1}{2})}{sKm\frac{1}{2}} \delta(\mathbf{\Delta k}-Km),
	\end{aligned}
\end{equation}

At this point if we consider a certain shape function $T_s(\mathbf{r}')$ that can be taken as a rectangular function with the size L
\begin{equation}\label{inter18}
	T_s(\mathbf{r}')=rect\bigg(\frac{\mathbf{r}'}{L}\bigg),
\end{equation}
according to  Eqs. \eqref{inter12},\eqref{inter13},\eqref{inter15aa} we will get the following expression for the diffracted wave 
\begin{equation}\label{inter19}
	\begin{aligned}
		E_{d}(\mathbf{\Delta k},\omega)& =\frac{k^2}{4\pi}\frac{e^{ikr}}{r}\frac{s}{d} \sum_{m=-\infty}^{\infty} \frac{sin(sKm\frac{1}{2})}{sKm\frac{1}{2}} \int rect\bigg(\frac{\mathbf{r}'}{L}\bigg) e^{i(Km- \mathbf{\Delta k})\mathbf{r}'} d\mathbf{r}' \\ 
		&=\frac{k^2}{4\pi}\frac{e^{ikr}}{r}\frac{sL}{d}\sum_{m=-\infty}^{\infty} \frac{sin(sKm\frac{1}{2})}{sKm\frac{1}{2}}\frac{sin[L(\mathbf{\Delta k}-Km)/2]}{L(\mathbf{\Delta k}-Km)/2}.
	\end{aligned}
\end{equation} 
For demonstrative purposes an alternate form can be written in the following way
\begin{equation}\label{inter20}
	\begin{aligned}
		E_{d}(\mathbf{\Delta k},\omega)& =\frac{k^2}{4\pi}\frac{e^{ikr}}{r}\frac{1}{d} \int\sum_{m=-\infty}^{\infty} \frac{sin(sKm\frac{1}{2})}{Km\frac{1}{2}} rect\bigg(\frac{\mathbf{r}'}{L}\bigg) e^{iKm\mathbf{r}'} e^{-i\mathbf{\Delta k}\mathbf{r}'} d\mathbf{r}' \\ 
		&=\frac{k^2}{4\pi}\frac{e^{ikr}}{r}\frac{1}{d}FT\big[\sum_{m=-\infty}^{\infty} \frac{sin(sKm\frac{1}{2})}{Km\frac{1}{2}} e^{iKm\mathbf{r}'} rect\bigg(\frac{\mathbf{r}'}{L}\bigg)\big]_{\mathbf{\Delta k}} \\
		&=\frac{k^2}{4\pi}\frac{e^{ikr}}{r}\frac{1}{d}FT\big[\sum_{m=-\infty}^{\infty} \frac{sin(sKm\frac{1}{2})}{Km\frac{1}{2}} e^{iKm\mathbf{r}'}\big]_{\mathbf{\Delta k}}*FT\big[rect\bigg(\frac{\mathbf{r}'}{L}\bigg)\big]_{\mathbf{\Delta k}}\\
		&=\frac{k^2}{4\pi}\frac{e^{ikr}}{r}\frac{sL}{d} \big[\sum_{m=-\infty}^{\infty} \frac{sin(s\mathbf{\Delta k}\frac{1}{2})}{s\mathbf{\Delta k}\frac{1}{2}} \delta(\mathbf{\Delta k}-Km)\big]*\big[\frac{sin[L(\mathbf{\Delta k})\frac{1}{2}]}{L(\mathbf{\Delta k})\frac{1}{2}}\big].
	\end{aligned}
\end{equation} 
In this way final expresion in Eq. \eqref{inter20} shows that diffracted wave $E_{d}(\mathbf{\Delta k},\omega)$ is just a convolution of Dirac comb function $\delta(\mathbf{\Delta k}-Km)$ modulated by the FT function of one grating period $sinc(s\mathbf{\Delta k}/2)$ with a FT from illumination function of the grating $sinc[L(\mathbf{\Delta k})/2]$. 
\begin{figure}[tb]
	\centering
	\includegraphics[width=1.0\textwidth]{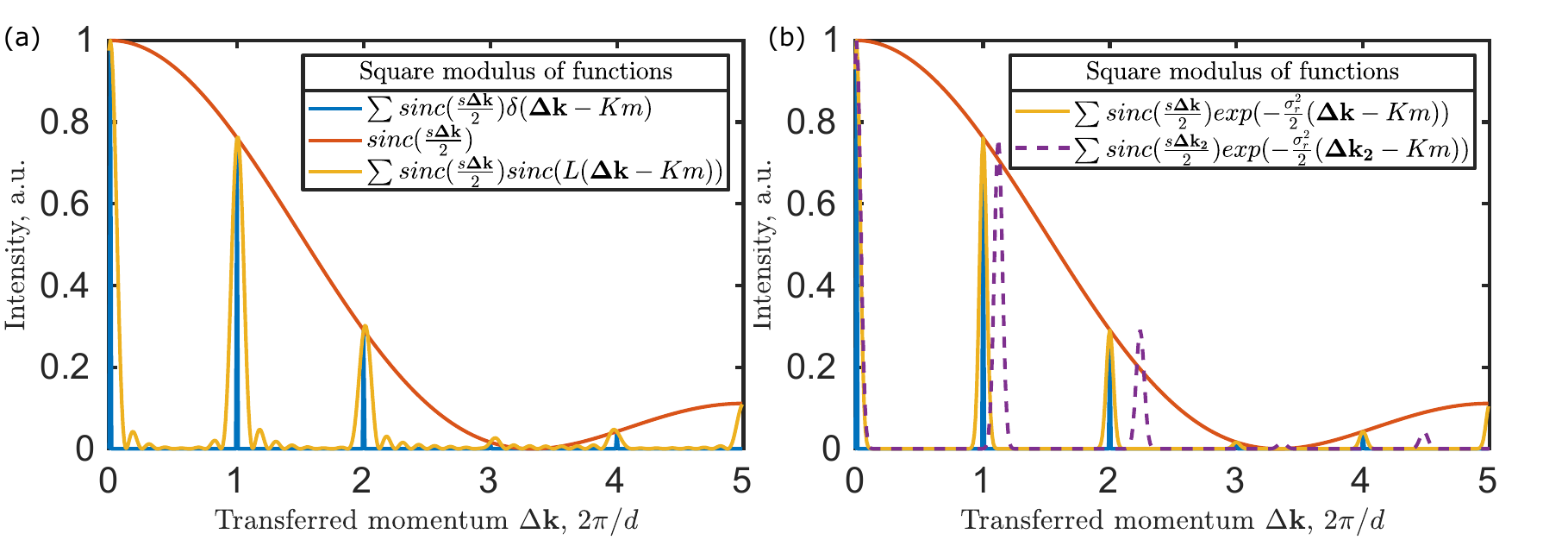}
	\caption{(a) A beam propagating through a grating with a period D and size of the period s. (b) Various functions for the diffracted field.} \label{FigG3}
\end{figure}As such one can use simple functions convinient for our next calculations. For example if one chooses in Eq. \eqref{inter8} illumination function $E_0(\mathbf{r})$ as a Gaussian function, 
\begin{equation}\label{inter21}
	E_0(\mathbf{r})=exp\bigg(-\frac{|\mathbf{r}|^2}{2\sigma_r^2}\bigg),
\end{equation}
under the condition that L $\gg$ $\sigma_r$ $\gg$ $\sigma_s$, the diffracted wave will be the convolution of a modulated periodic Delta function with a Gaussian function 
\begin{equation}\label{inter22}
	\begin{aligned}
		E_{d}(\mathbf{\Delta k},\omega) = 
		\frac{k^2}{4\pi}\frac{e^{ikr}}{r}\frac{\sqrt{2\pi \sigma_r}s}{d}\sum_{m=-\infty}^{\infty} \frac{sin(sKm\frac{1}{2})}{sKm\frac{1}{2}}exp\bigg(\frac{-\sigma_r^2(\mathbf{\Delta k}-Km)^2}{2}\bigg).
	\end{aligned}
\end{equation} 
Expression in Eq. \eqref{inter20} and \eqref{inter22} is shown in Fig. \ref{FigG3}. At this point we may closer consider only first order $m$=1 in the dispersion $x$-direction for the situation shown in Fig. \ref{FigG0}(b), assuming $\theta_I=0$
\begin{equation}\label{inter24}
	\begin{aligned}
		E_{d}({k}_x,\omega) = 
		\frac{k^2}{4\pi}\frac{e^{ikr}}{r}\frac{\sqrt{2\pi \sigma_r}s}{d} \frac{sin(sKm\frac{1}{2})}{sKm\frac{1}{2}}exp\bigg(\frac{-\sigma_r^2({k}_x-K)^2}{2}\bigg).
	\end{aligned}
\end{equation} 
In this case diffracted wave is Gaussian function in the (${k}_x,\omega$) $-$ domain shifted by the spatial wave number $K$. 

Now let us assume that the incident wave has a tiny shift in energy $k_{0}'=k_0-\Delta k_0$ from the resosant wave with $k_0$ such that $\Delta\omega \ll \omega_0$, but the incident angle is kept the same $\theta_I=0$. Then for the diffracted amplitude we have
\begin{equation}\label{inter25}
	\begin{aligned}
		E_{d}(k_x,\Delta\omega) = 
		E_0\frac{\sqrt{2\pi \sigma_r}s}{d} \frac{sin(sK\frac{1}{2})}{sK\frac{1}{2}}exp\bigg(\frac{-\sigma_r^2}{2}(k_x -[\Delta k_{0x}+K])^2\bigg),
	\end{aligned}
\end{equation} 
where $k_x=k_0sin(\theta_{D})$ and $\Delta k_{0x}=\Delta k_{0}sin(\theta_{D})$.
In this case one can see from the Eq. \eqref{inter25} that amplitude in the first diffraction order is a Gaussian function in ($k_x,\Delta\omega$)-domain, which is shifted from the resonant component $k_{0x}=k_0sin(\theta_{I})=0$ by the $\Delta k_{0x}+K$ (see for example Fig. \ref{FigG3}(b)). 
Now, for simplicity, we may shift our origin of $k_x$-axis to the position of the first order, i. e. by K, so that in the new system a diffracted wave with $k_0$ and the wave with the energy shift $\Delta k_0$ are
\begin{equation}\label{inter25a}
	\begin{aligned}
		&E_{1}(k_x,\omega) \sim 
		exp\bigg(\frac{-\sigma_r^2}{2}(k_x)^2\bigg) \\
		&E_{2}(k_x,\Delta\omega) \sim 
		exp\bigg(\frac{-\sigma_r^2}{2}(k_x -\Delta k_{0x})^2\bigg) \\
	\end{aligned}
\end{equation} 
At this point we have to emphasize, although it is common and convinient to use  $\{k_x,\omega\}$-domain, in our case it will be interesting to work in  $\{x,\omega\}$-domain. We will define the Spatial-frequency Fourier domain transform as follows
\begin{align}
	&\hat{E}(k_x, \omega ) =  \frac{1}{2\pi} \int_{-\infty}^{\infty} dx~ \bar{E}(x,\omega)\exp(-ik_x x) ~.
	\cr &
	\bar{E}(x,\omega) =  \int_{-\infty}^{\infty} dk_x~\hat{E}(k_x,\omega ) \exp(ik_x x)~.
	\label{inter26}
\end{align}
According to Eq.\eqref{inter26} 
\begin{equation}\label{inter27}
	\begin{aligned}
		\bar{E_2}(x,\omega)&= \int_{-\infty}^{\infty} dk_x~E_2(k_x-\Delta k_{0x},\omega ) \exp(ik_x x)\\
		&=[\text{substitution}~ q_x=k_x-\Delta k_{0x};~ dq_x=dk_x;~ k_x=q_x+\Delta k_{0x}]\\
		&=\int_{-\infty}^{\infty} dq_x~E_2(q_x,\omega ) \exp(i[q_x+\Delta k_{0x}]x)\\
		&=\exp(i\Delta k_{0x}x)\int_{-\infty}^{\infty} dq_x~E_2(q_x,\omega )\exp(iq_x x)\\
		&=\exp(i\Delta k_{0x}x)\bar{E_1}(x,\omega),
	\end{aligned}
\end{equation} 
which is basically a Shift theorem. As the result one can see that the angular dispersion in the $(k_x,\omega)$ domain results in the additional space-dependent phase in the $(x,\omega)$-domain.
\begin{equation}\label{inter27a}
	\begin{aligned}
		\bar{E_2}(x,\omega)= 
		\exp(i\Delta k_{0x}x)\bar{E_1}(x,\omega),
	\end{aligned}
\end{equation} 
At this point, taking into account geometry shown in Fig. \ref{FigG0} and Eq. \ref{inter17a}  we may introduce a dispersion parameter $p$ as
\begin{equation}\label{inter28}	
	p = \frac{d k_{x}}{d \omega}=\frac{k d \theta_d}{d \omega}=\frac{\lambda}{c}\frac{d \theta_d}{d \lambda}=\frac{\lambda }{c }\frac{\bar{d}}{cos(\theta_d)},
\end{equation} 
where $\bar{d}$=$1/d$ is the line density. In this case, the diffracted wave $E_2(x,\omega_0+\Delta \omega)$ shifted by  $\Delta \omega$ from the resonant $E_1(x,\omega_0)$ satisfies the following equation in the $(x,\omega)$-domain
\begin{equation}\label{inter29}
	\begin{aligned}
		E(x,\omega_0+\Delta \omega)= \exp(ip\Delta \omega x)E(x,\omega_0),
	\end{aligned}
\end{equation}
The phasor in the Eq. \eqref{inter29} can be considered physically as the factor which determines planes of the phase fronts for each particular monochromatic component $\omega$ in the $\{\mathbf{r},\omega\}$ domain. We have to note here although the $\{\mathbf{k},\omega\}$ domain is generally accepted and commonly used, the $\{\mathbf{r},\omega\}$ domain is more adequate for our further statistical analysis.
\subsection{\label{sec:refl} Dispersion in the reflection geometry}
\begin{figure}[tb]
	\centering
	\includegraphics[width=1.0\textwidth]{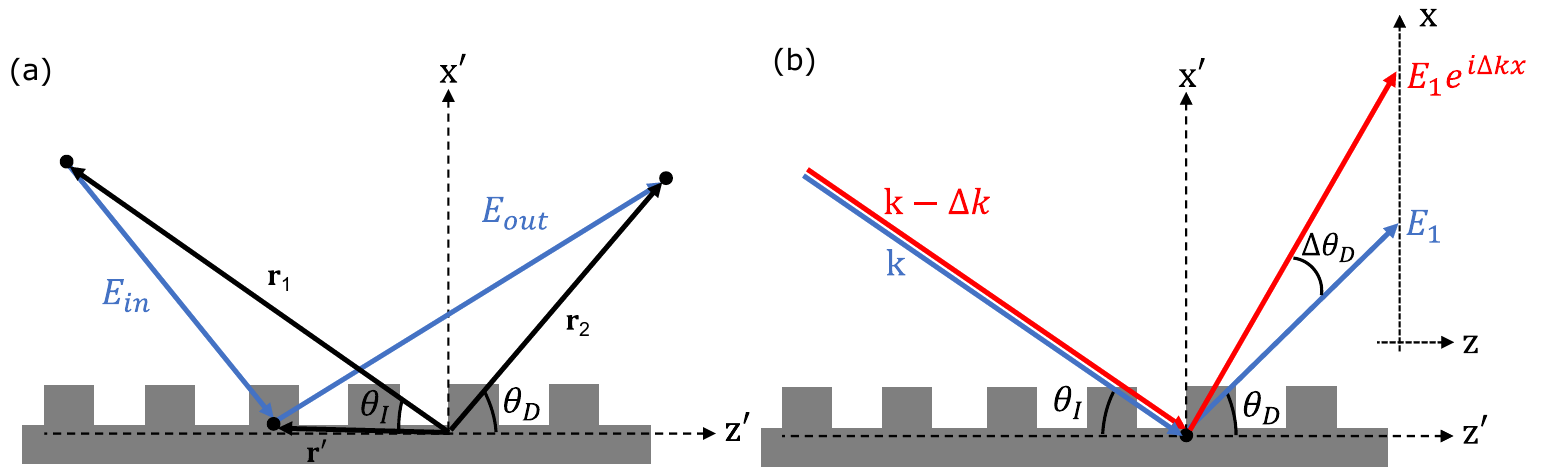}
	\caption{(a) Beam reflection scheme. (b) A beam propagating through a grating monochromator undergoes a phase shift according to grating specifications.} \label{FigG1}
\end{figure}
In this section we would like to consider a diffraction grating in the reflection geometry (see Fig. \ref{FigG1}). We assume that penetration of the incoming field $E_{in}$ into the surface of the grating is neglectable ($x'^2 \ll y'^2+z'^2$). For simplicity the paraxial approximation is used as well, implying that $\mathbf{r'} \ll \mathbf{r_1}$ and  $\mathbf{r'} \ll \mathbf{r_2}$. In this case the grazing incidence $\theta_I$ for $E_{in}$ is the same angle between axis $z'$ and $\mathbf{r_1}$ (see Fig. \ref{FigG1}). Similar applies to the angle of reflection. Considering this picture, we may take in Eq. \eqref{inter8} the Green function as
\begin{equation}\label{ref1}
	G(\mathbf{r_2}-\mathbf{r'},\omega) = \frac{1}{4\pi}\frac{e^{ik_0|\mathbf{r_2}-\mathbf{r}'|}}{|\mathbf{r_2}-\mathbf{r}'|},
\end{equation}
and incoming spherical wave as
\begin{equation}\label{ref2}
	E_{in}=E_{0}(\mathbf{r'}-\mathbf{r_1},\omega) = \frac{e^{ik_0|\mathbf{r}'-\mathbf{r_1}|}}{|\mathbf{r}'-\mathbf{r_1}|}.
\end{equation}
In this case taking into account the geometry shown in Fig. \ref{FigG1}, Eq. \eqref{inter14a} for the periodic surface in $z$-direction, far-filed approxiation, additionaly changing the order of summation and integration, the diffracted wave is
\begin{equation}\label{ref3}
	\begin{aligned}
		E_d(\mathbf{r_1},\mathbf{r_2},\omega) &= \frac{k^2}{4\pi}\frac{1}{d}\sum_{m=-\infty}^{\infty}\int \frac{e^{ik_0|\mathbf{r_2}-\mathbf{r}'|}}{|\mathbf{r_2}-\mathbf{r}'|}e^{iKmz'}\frac{e^{ik_0|\mathbf{r}'-\mathbf{r_1}|}}{|\mathbf{r}'-\mathbf{r_1}|}d\mathbf{r}'\\
		&=E_0\sum_{m=-\infty}^{\infty}\int e^{ik_0\big[|\mathbf{r_2}|-\frac{\mathbf{r_2}\mathbf{r}'}{|\mathbf{r_2}|}\big]}e^{ik_0\big[|\mathbf{r_1}|-\frac{\mathbf{r_1}\mathbf{r}'}{|\mathbf{r_1}|}\big]}e^{iKmz'}d\mathbf{r}'\\
		&=E_0e^{ik_0(r_2+r_1)}\sum_{m=-\infty}^{\infty}\int
		e^{-ik_0(\mathbf{n_2}\mathbf{r}'+\mathbf{n_1}\mathbf{r}')}e^{iKmz'}d\mathbf{r}',
	\end{aligned}
\end{equation}
where $\mathbf{n_2}$ and $\mathbf{n_1}$ are the unit vectors in the directions of $\mathbf{r_2}$ and $\mathbf{r_1}$ accordingly.
The integral of Eq. \eqref{ref3} over $z'$ is the delta function,
\begin{equation}\label{ref3a}
	E_{z}\sim \sum_{m=-\infty}^{\infty}\int e^{i(Km-\mathbf{\Delta k})z'}d{z}' = \sum_{m=-\infty}^{\infty} \delta(Km-\mathbf{\Delta k}),
\end{equation}
which again defines a dispersion condition for the reflected wave 
\begin{equation}\label{ref3b}
	\mathbf{\Delta k}=k_0(\mathbf{n_2}+\mathbf{n_1})=Km.
\end{equation}
Acounting for the geometry shown in Fig. \ref{FigG1} the dispersion condition is
\begin{equation}\label{ref3c}
	cos~\theta_{D} - cos~\theta_I   = \frac{\lambda}{d}m
\end{equation}
Using mentioned above approximations, we may also expand the term in the phase of incident wave in Eq. \ref{ref2} as
\begin{equation}\label{ref4}
	\begin{aligned}
		|\mathbf{r}'-\mathbf{r_1}|& = \sqrt{(r_1 sin ~\theta_I)^2+y'^2+(r_1 cos ~\theta_I+z')^2}  \\
		&=\sqrt{(r_1 sin ~\theta_I)^2+y'^2+(r_1cos~\theta_I)^2 +2r_1 z'cos ~\theta_I + z'^2}\\
		&=\sqrt{r_1^2+y'^2+2r_1 z'cos ~\theta_I + z'^2} \\
		&=\sqrt{r_1^2+2r_1cos~\theta_I+(z'cos~\theta_I)^2+y'^2+(z'sin~\theta_I)^2}\\
		&=\sqrt{(r_1+z'cos~\theta_I)^2+y'^2+(z'sin~\theta_I)^2}\\
		&\approx(r_1+z'cos~\theta_I)\big[1+ \frac{y'^2}{2(r_1+z'cos~\theta_I)^2}\frac{(z'sin~\theta_I)^2}{2(r_1+z'cos~\theta_I)^2}  \big]\\
		&=r_1+z'cos~\theta_I + \frac{y'^2}{2r_1(1+\frac{z'cos~\theta_I}{r_1})}+\frac{(z'sin~\theta_I)^2}{2r_1(1+\frac{z'cos~\theta_I}{r_1})}\\
		&\approx r_1+z'cos~\theta_I +\frac{y'^2}{2r_1}\big(1-\frac{z'cos~\theta_I}{r_1}\big)+\frac{(z'sin~\theta_I)^2}{2r_1}\big(1-\frac{z'cos~\theta_I}{r_1}\big)\\
		&=r_1+z'cos~\theta_I + \frac{y'^2}{2r_1} + \frac{(z'sin~\theta_I)^2}{2r_1} - \frac{z'^3 sin^2~\theta_I~cos~\theta_I}{2r_1^2}-\frac{y'^2z'cos~\theta_I}{2r_1^2}.
	\end{aligned}
\end{equation}
By the analogy the term in the Green function (Eq. \eqref{ref1}) can be expanded as
\begin{equation}\label{ref5}
	\begin{aligned}
		|\mathbf{r_2}-\mathbf{r}'| =\sqrt{(r_2 sin ~\theta_D)^2+y'^2+(r_2 cos ~\theta_D-z')^2} \approx\\
		r_2-z'cos~\theta_D + \frac{y'^2}{2r_2} + \frac{(z'sin~\theta_D)^2}{2r_2} + \frac{z'^3 sin^2~\theta_D~cos~\theta_D}{2r_2^2}+\frac{y'^2z'cos~\theta_D}{2r_2^2}.
	\end{aligned}
\end{equation}
The total phase under the integral of the diffracted wave in Eq. \eqref{ref3} is usually written as power series
\begin{equation}\label{ref6}
	\phi=ik_0(F_{00}+F_{10}z'+F_{20}z'^2+F_{02}y'^2+F_{12}z'y'^2+F_{30}z'^3),
\end{equation}
where coefficients defined as follows

\begin{align}\label{ref7}
	&F_{00} = r_1+r_2   ~,
	\cr &
	F_{10} = \frac{Km}{k_0} + cos~\theta_I - cos~\theta_{D} ~,
	\cr &
	F_{20} = \frac{1}{2}\big[\frac{sin^2~\theta_I}{r_1}  + \frac{sin^2~\theta_D}{r_2}\big] ~,
	\cr &
	F_{02} = \frac{1}{2}\big[\frac{1}{r_1}+\frac{1}{r_2}\big]~,
	\cr &
	F_{12} =\frac{1}{2}\big[-\frac{cos~\theta_I}{r_1^2}  + \frac{cos~\theta_D}{r_2^2}\big] ~,
	\cr &
	F_{30} = \frac{1}{2}\big[-\frac{sin^2~\theta_I ~cos~\theta_I}{r_1^2}  + \frac{sin^2~\theta_D~cos~\theta_{D}}{r_2^2}\big] ~.
\end{align}

In this case we got additional coefficients $F_{20} - F_{30}$ describing different type of aberrations. Coefficients $F_{20}$ and $F_{02}$ describing defocusing, $F_{12}$ describes the astigmatic aberation and $F_{30}$ describes the coma. Effect of aberrations on the properties o the beam will be given in next chapters. In order to obtain non-zero contribution of the integral in Eq. \eqref{ref3} for the diffracted wave $E_d(\mathbf{r_1},\mathbf{r_2},\omega)$, coefficients in Eqs. \eqref{ref7} should be equal to zero or in other words aberrations should be reduced to minimum. The condition $F_{10}=0$ implies the same angular distribution law, as in Eq. \eqref{ref3c}

We are mostly interested in vertical dispersive direction, and projections of $k$-vector on $x$-axis. In this case in the paraxial approximation ($sin ~\theta_{D} \approx \theta_{D}$) acounting for our geometry shown in Fig. \ref{FigG1}(a)
\begin{equation}\label{ref9}
	\frac{d \theta_d}{d \lambda}=-\frac{m}{\theta_{D} d}.
\end{equation}
For us it is interesting to trace the change of $k$-vector in the dispersion direction acounting for our geometry, namely dispersion $dk_x/d\omega$ at fixed $z$ position in the first order
\begin{equation}\label{ref10}	
	p = \frac{d k_{x}}{d \omega}=-\frac{\lambda }{c}\frac{\bar{d}}{\theta_D},
\end{equation} 
where negative sign indicates on contraction of the projection $k$ on the x-axis upon increasing the energy of the photon beam from the resosnant. The opposite behaviour is seen for the $-1$ order. By the analogy with Eq. \eqref{inter29} we may write 
\begin{equation}\label{ref11}
	\begin{aligned}
		E_2(x,\omega_0+\Delta \omega)= \exp(-i\frac{\lambda \bar{d} \Delta \omega}{c \theta_D })E_1(x,\omega_0).
	\end{aligned}
\end{equation}

For efficient monochromisation, the photon beam diffracted by the grating is focused into the plane of the exit slit aperture. Focusing can be achieved by using an additional focusing element after a plane grating or by using a self-focusing grating such as a Variable Line Spacing (VLS) grating. The calculation of the photon beam diffracted by a VLS grating is carried out by taking the groove spacing $d$ as a function of spatial coordinate $x'$ along the grating (Eq. \eqref{inter14})

\begin{equation}\label{vls1}
	d(x')=d_0+d_1x'+d_2x'^2+...,
\end{equation}

where $d_0$ is the groove spacing at the center of the grating, and $d_1$ and $d_2$ are the VLS coefficients for the ruled width variation along the $x'$ direction. Such a periodic structure allows the photon beam to be spectrally dispersed and simultaneously focused onto the exit slit plane. 

In the case of a VLS grating, the focusing term of the optical path function (see Eq. \eqref{ref6}) can be written as follows, taking into account Eqs. \eqref{vls1}, \eqref{inter14}, \eqref{ref4}, and \eqref{ref5}

\begin{equation}\label{vls2}
	F_{20} = \frac{1}{2}\big[\frac{sin^2~\theta_I}{r_1}  + \frac{sin^2~\theta_D}{r_2}\big] + \frac{\lambda m}{d_1}.
\end{equation}

where the focus condition is $F_{20}=0$. Assuming 1:1 focusing, the focal distance of the VLS grating is given by

\begin{equation}\label{vls3}
	f_{g} = \frac{sin^2 ~\theta_{D}+sin^2~\theta_{I}}{2\lambda m \bar{d_1}}.
\end{equation}

Note that the VLS coefficient $\bar{d_1}$ can be negative. 

\subsection{Resolution under Rayleigh criterion}\label{Rayleigh}
In the first case, we have to look closely at Eq. \eqref{inter19} and Fig.\ref{FigG3}(a). Primary peaks of the intensity distribution function $|E_{d}(\mathbf{\Delta k},\omega)|^2$ corresponding to various diffraction orders have a width inversely proportional to the size of the grating L. Additionally to that, secondary peaks are present in the distribution between primary maxima. The Maxima of the secondary peaks are separated by the zero intensity points at 
\begin{equation}\label{es3a}
	L(\mathbf{\Delta k}-Km)/2 =  n\pi,
\end{equation} 
where $n=\pm 1, \pm 2 ...$. Replacing the size of the grating by the number of illuminated grooves $N_g$ and the groove period $d$, considering the first order beam shifted to K, the condition is satisfied when
\begin{equation}\label{es3b}
	sin(\theta_{D})-sin(\theta_I)=\frac{n\lambda}{N_gd}.
\end{equation}
The difference between two maxima of the secondary peaks is
\begin{equation}\label{es3c}
	\Delta=\frac{\lambda}{N_gd}.
\end{equation}
According to Eq.\eqref{inter17a}, a shift of the wavelength by $\Delta \lambda$ in the first order ($m=1$) will change the maximum of the primary peak by 
\begin{equation}\label{es3d}
	\Delta'=\frac{\Delta\lambda}{d}.
\end{equation}
The Rayleigh criterion implies that two components of the intensity distribution may be resolved if the primary maximum of the first component (i.e., intensity peak of the shifted wave by $\Delta \lambda$) coincides with the first minimum of the second component (i.e., with the secondary peak) ($\Delta'$=$\Delta$). The latter criterion, according to Eqs. \eqref{es3c} and \eqref{es3d} defines the resolution of the monochromator
\begin{equation}\label{es3e}
	\frac{\Delta \lambda}{\lambda}=\frac{1}{N_g}.
\end{equation}
Thus, one can see that under the Rayleigh criterion the resolution of the monochromator is determined by the number of illuminated grooves, which is equivalent to the case of fully close exit slit of the monochromator. 

\subsection{Spatial correlation function}\label{sec:SpatCorSupp}

Since for us it is convinient to work in the $\{\mathbf{r},\omega\}$ domain, acounting that Gaussian process $E(\mathbf{r},t)$ in time domain linked to $E(\mathbf{r},\omega)$ by Fourier transform, the corresponding correlation function in frequency domain is
\begin{equation} \label{coh2}
	\Gamma_{\omega} (\mathbf{r_1},\mathbf{r_2},\omega_1,\omega_2)=<E^*(\mathbf{r_1},\omega_1)E(\mathbf{r_2},\omega_2)>.	
\end{equation}
Taking into account mathematical description of the source (\ref{sourceM}) one can obtain spatial correlation function, starting with second-order field correlation function in the frequency domain 
\begin{equation} \label{coh4}
	\Gamma_{\omega} (\mathbf{r_1},\mathbf{r_2},\omega_1,\omega_2)=<\sum_{k=1}^{N_e}\sum_{m=1}^{N_e} E^*_{k\perp}(\boldsymbol{r_1},\omega_1)E_{m\perp}(\boldsymbol{r_2},\omega_2)e^{-i\omega_1 t_k+i\omega_2 t_m}>.
\end{equation}
We can also split the ensemble sum into two parts
\begin{equation} \label{coh5}
	<\sum_{k=1}^{N_e}\sum_{m=1}^{N_e}> = <\sum_{k=m=1}^{N_e}>+<\sum_{k\neq m}>
\end{equation}
The first ensemble sum $<\sum_{k=m=1}^{N_e}>$ of Eq. \eqref{coh5} implies correlation of each individual electron with itself. The second ensemble sum  $<\sum_{k\neq m}>$ implies correlations between different electrons.
However, the effects arising in the second case 
are hardly of particular interest at the synchrotron facilities, since radiation wavelengths are much shorter than the bunch length. In this regard, we omit the second sum in the Eq. \eqref{coh5}, taking into account Eq. \eqref{undi45} and the fact that the average of the product of two independently distributed random realizations is the product of their individual averages we may write
\begin{equation} \label{coh6}
	\Gamma_{\omega} (\mathbf{r_1},\mathbf{r_2},\omega_1,\omega_2)=\sum_{k=1}^{N_e}<e^{i(\omega_2-\omega_1)t_k}>_{t_k}< E^*_{k\perp}(\boldsymbol{r_1},\omega_1)E_{k\perp}(\boldsymbol{r_2},\omega_2)>_{\boldsymbol{\eta},\mathbf{l},\gamma}
\end{equation}
The ensemble average of the first term according to definition in Eq. \eqref{undi6a}  is
\begin{equation} \label{coh7}
	<e^{i(\omega_2-\omega_1)t_k}>_{t_k} = \int e^{i(\omega_2-\omega_1)t_k}f(t_k)dt_k=G_{\omega}(\omega_2-\omega_1),
\end{equation}
where $f(t_k)$ is the bunch pulse profile, and $G_{\omega}(\omega_2-\omega_1)$ its Fourier transform or spectral correlation function. 

Taking into account the previous assumption in \ref{PulseSR} that the monochromator can not resolve a single spike in spectrum, together with the analysis of characteristic scales, under the condition that the functions $E_{\perp}(\boldsymbol{r},\omega)$  and $G_{\perp}(\mathbf{r_1},\mathbf{r_2},\omega)$ on the scale $\Delta\omega/\omega$ vary much slower than $G_{\omega}(\omega_2-\omega_1)$ (see Fig. \ref{Fluc1}) and $\Delta\omega\gg 1/\sigma_t$  one can substitute
spectral correlation function $G_{\omega}(\omega_2-\omega_1)$ with Dirac $\delta(\omega_2-\omega_1)$ function implying that no correlations occur between different monochromatic components of the synchrotron radiation in this quasi-stationary case \cite{geloni2008transverse}.
 Therefore 
Eq. \eqref{coh7} simplifies to
\begin{equation} \label{coh8}
	\Gamma_{\omega} (\mathbf{r_1},\mathbf{r_2},\omega)=N_e \delta(\omega_2-\omega_1)G_{\perp}(\mathbf{r_1},\mathbf{r_2},\omega),
\end{equation}
where $G_{\perp}(\mathbf{r_1},\mathbf{r_2},\omega)$ is 
\begin{equation} \label{coh9}
	G_{\perp}(\mathbf{r_1},\mathbf{r_2},\omega)=<E^*_{\perp}(\boldsymbol{r_1},\omega)E_{\perp}(\boldsymbol{r_2},\omega)>_{\boldsymbol{\eta},\mathbf{l},\gamma}.
\end{equation}
We see from Eq. \eqref{coh8} that correlation in $\{\mathbf{r},\omega\}$ domain is factorized into separate spatial and spectral parts. However, radiation fields $E_{\perp}(\boldsymbol{r},\Delta\omega)$ become spatially dependent from the frequency offset $\Delta\omega$ after interaction with the grating. 
Upon following definitions 

\begin{eqnarray}
	&& x_1 = \bar{x} + \frac{\Delta x}{2} \cr
	&& x_2 = \bar{x} - \frac{\Delta x}{2} ~.
	\label{moi2}
\end{eqnarray}
we can define in full generality the spatial correlation function of the electric field between points $x_1$ and $x_2$ as

%
%

\begin{equation}
	G_{\perp}(\bar{x},\Delta x) = \frac{1}{2\pi}   \int_{-\infty}^{\infty} d \Delta\omega ~ {E}^*\left(\bar{x} + \frac{\Delta x}{2},\Delta\omega\right){E}\left(\bar{x} - \frac{\Delta x}{2},\Delta\omega\right)~.
	\label{moi3_01}
\end{equation}
Degree of transverse coherence is given accordingly
\begin{eqnarray}
	\zeta = \frac{\int_{-\infty}^{\infty}\int_{-\infty}^{\infty} d\bar{x}~ d\Delta x~ \left|G_{\perp}(\bar{x},\Delta x)\right|^2}{\left|\int_{-\infty}^{\infty} d\bar{x} ~G_{\perp}(\bar{x},\Delta x=0) \right|^2} ~.
	\label{moi3a_01}
\end{eqnarray}
Here one can also define the spatial (double) Fourier-transform of $G_{\perp}$ with respect to $x_1$ and $x_2$, which is a function of $k_{x1}$ and $k_{x2}$ as

\begin{align}
	G_{\perp}(k_{x1},k_{x2})  =& \int_{-\infty}^{\infty} d x_1 \int_{-\infty}^{\infty} d x_2 G_{\perp}(x_1,x_2) \exp(+i k_{x1} x_1 - i k_{x2} x_2) \cr =&   \frac{1}{2\pi} \int_{-\infty}^{\infty} d \Delta\omega {E}^*\left(\Delta\omega, k_{x1}\right){E}\left(\Delta\omega,k_{x2}\right)\cr &
	\label{moi4}
\end{align}
where one may also legitimately introduce the additional notation

\begin{eqnarray}
	&& k_{x1} = \bar{k}_{x} + \frac{\Delta k_{x}}{2} \cr
	&& k_{x2} = \bar{k}_{x} - \frac{\Delta k_{x}}{2} ~.
	\label{x1x2}
\end{eqnarray}


\bibliography{sample2}

\end{document}